\documentclass[12pt]{article}
\usepackage{epsfig,amssymb}
\usepackage{latexsym}

\newcommand{\bq}{\begin{equation}}
\newcommand{\eq}{\end{equation}}
\newcommand{\bqa}{\begin{eqnarray}}
\newcommand{\eqa}{\end{eqnarray}}
\newcommand{\ben}{\begin{enumerate}}
\newcommand{\een}{\end{enumerate}}
\newcommand{\bc}{\begin{center}}
\newcommand{\ec}{\end{center}}
\newcommand{\bqb}{\begin{eqnarray*}}
\newcommand{\eqb}{\end{eqnarray*}}

\def\sw{s_W}
\def\cw{c_W}

\def\mz{m_Z}
\def\tchi{\tilde \chi}

\def\ssf{s_{\tilde \theta_f}}
\def\csf{c_{\tilde \theta_f}}

\def\pr#1#2#3{ Phys. Rev. ${\bf{#1}}$, #2 (#3)}
\def\prl#1#2#3{ Phys. Rev. Lett. ${\bf{#1}}$, #2 (#3)}
\def\pl#1#2#3{ Phys. Lett. ${\bf{#1}}$, #2 (#3)}
\def\prep#1#2#3{ Phys. Rep. ${\bf{#1}}$, #2 (#3)}

\def\np#1#2#3{ Nucl. Phys. ${\bf{#1}}$, #2 (#3)}

\def\epj#1#2#3{ Eur. Phys. J. ${\bf{#1}}$, #2 (#3)}
\def\ijmp#1#2#3{ Int. J. Mod. Phys. ${\bf{#1}}$, #2 (#3)}

\def\fortp#1#2#3{Fortsch. Phys. ${\bf{#1}}$, #2 (#3)}
\def\jphys#1#2#3{J. Phys. ${\bf{#1}}$, #2 (#3)}

\def\aop#1#2#3{Annals of Phys. ${\bf{#1}}$, #2 (#3)}

\def\polon#1#2#3{Acta Phys. Polon. ${\bf{#1}}$, #2 (#3) }

\begin{document}
\pagenumbering{arabic}
\thispagestyle{empty}
\def\thefootnote{\fnsymbol{footnote}}
\setcounter{footnote}{1}

\begin{flushright}
July 17, 2012\\
arXiv:1205.4547 [hep-ph]\\
 \end{flushright}

\vspace{2cm}

\begin{center}
{\Large {\bf A  supersimple analysis of $e^-e^+\to  t \bar t$ at high energy}}.\\
 \vspace{1cm}
{\large G.J. Gounaris$^a$ and F.M. Renard$^b$}\\
\vspace{0.2cm}
$^a$Department of Theoretical Physics, Aristotle
University of Thessaloniki,\\
Gr-54124, Thessaloniki, Greece.\\
\vspace{0.2cm}
$^b$Laboratoire Univers et Particules de Montpellier,
UMR 5299\\
Universit\'{e} Montpellier II, Place Eug\`{e}ne Bataillon CC072\\
 F-34095 Montpellier Cedex 5.\\
\end{center}

\vspace*{1.cm}
\begin{center}
{\bf Abstract}
\end{center}

According to supersimplicity in MSSM,
a renormalization scheme (SRS) may be defined for any high energy 2-to-2 process,
to the 1loop EW order;  where  the helicity conserving (HC) amplitudes,
 are expressed as a linear combination of just three universal logarithm-involving forms.
  All other helicity  amplitudes  vanish asymptotically.
Including to these SRS amplitudes the corresponding counter terms,
the "supersimple" expressions for the high energy HC amplitudes, renormalized
on-shell,  are obtained.

Previously,  this property was  noted for a large number of processes that
do  not involve   Yukawa interactions   or renormalization group  corrections.
Here we extend it   to  $e^-e^+\to  t \bar t$,
which does involve large  Yukawa and renormalization
group contributions. We  show that the resulting  "supersimple"
expressions may provide an accurate description, even  at energies
comparable to  the SUSY scale. Such descriptions clearly
identify the origin of the important SUSY effects, and  they may be used for  quickly
 constraining  physics contributions, beyond MSSM.

\vspace{0.5cm}

\def\thefootnote{\arabic{footnote}}
\setcounter{footnote}{0}
\clearpage

\section{ Introduction}

In a recent paper \cite{super}, we have shown that at the 1loop EW order of
several high energy 2-to-2 processes  in MSSM, a remarkably simple structure arises
for the helicity conserving\footnote{The definitions of the HC and HV amplitudes appear
in \cite{super} and are repeated below.} (HC)
amplitudes; which are the only surviving amplitudes  in this limit \cite{heli1,heli2}.
This structure, which  has been  called "supersimplicity",  involves just
three forms: two Sudakov-like forms, containing a log or a
log-squared function of the ratio of a  Mandelstam variable
with respect to masses, together  with an additional  energy independent term;
and a squared-log of the  ratio of two Mandelstam variables, to which  $\pi^2$  is added.

In \cite{super}, a supersimplicity renormalization scheme (SRS) was defined, where the
high energy HC amplitudes exactly have the above "supersimplicity"  structure,
without any additional term.
Adding to this "supersimplicity" amplitudes, some "residual" constant contributions,
which are viewed as counter terms (c.t.); the
"supersimple" expressions for the high energy  HC amplitudes in the on-shell  renormalization
scheme \cite{OS} are obtained.

Such "supersimple" results arise in MSSM after many  cancelations, among  much more complicated
  contributions, involving standard and supersymmetric particle exchanges.
While deriving them,  it is fascinating to observe
how the SUSY couplings conspire to achieve
the supersimple structure for the high energy on-shell HC amplitudes, and at the
same time to force the  helicity violating (HV) amplitudes to vanish \cite{super}.

In SM, where such conspiracies do not appear, additional linear logarithms
of  ratios of Mandelstam variables arising from boxes appear \cite{super}, which
 cannot be thought as a combination of Sudakov-like
 forms \cite{MSSMrules1, MSSMrules2, MSSMrules3, MSSMrules4}. Furthermore,
 additional residual constants are needed to describe the high energy (on-shell renormalized)
 HC amplitudes; while nothing is generally
known, for  the HV amplitudes.

The  supersimplicity structure was shown in \cite{super} for
a large number of MSSM processes,   which did not involve
any Yukawa terms or  renormalization group corrections,
to the  electroweak (EW) couplings. For $ug\to dW^+$ in particular,  the  supersimple high energy
 expressions for the HC amplitudes were considered in some  detail. Such expressions
 were  found  to  provide an accurate description,
 even at  energies comparable to the  SUSY scale  \cite{super}.\\

In the present work we extend the  analysis of \cite{super},
to a process involving
renormalization group contributions and large Yukawa terms.
Assuming that sometime in the
future  a high energy $e^-e^+$ collider (LC) will be built,
we consider the 1loop EW corrections to the process
\bq
e^-(l,\lambda)+e^+(l',\lambda') \to t(p,\mu)+\bar t(p',\mu')~~, \label{eett-process}
\eq
where $(l,l',p,p')$ denote the momenta, and
$(\lambda, \lambda', \mu, \mu') $ the helicities of the incoming and outgoing particles.
The corresponding helicity amplitudes,  denoted as
\bq
F(e^-e^+\to t\bar t)_{\lambda \lambda' \mu \mu'} ~~, \label{F-amplitude}
\eq
 are separated into two classes: the
 helicity conserving (HC) amplitudes  satisfying
\bq
\lambda +\lambda'= \mu + \mu' ~~; \label{heli-cons}
\eq
and the helicity violating  ones (HV), where (\ref{heli-cons}) is not respected.
Provided we ignore\footnote{As we have done also in \cite{super}.}
CP-violating couplings in MSSM, the amplitudes
(\ref{F-amplitude}) satisfy  \cite{JW, Chang, Dj}
\bq
F(e^-e^+\to t\bar t)_{\lambda, \lambda', \mu, \mu'}=
F(e^-e^+\to t\bar t)_{-\lambda',- \lambda, - \mu', -\mu} ~~.\label{CP-constraint}
\eq

Process (\ref{eett-process}),  indeed involves  large Yukawa interactions
affecting the final $t \bar t $ state; while  the existence   of
gauge boson self-energy contributions,  generates    renormalization group
(RG) logs and  large $\Delta\rho$-type terms\footnote{See Sect. 3.2}.
Our purpose is to investigate  how
supersimplicity  is affected by such     contributions.\\

Neglecting the electron mass, non-vanishing helicity amplitudes always satisfy
$\lambda+\lambda'=0$, which combined with (\ref{CP-constraint}), means that there exist
only two independent HV  amplitudes, for which  we take  $F_{-+--}, ~ F_{+---}$.
As discussed in connection to
Fig.\ref{HV-fig}, these HV amplitudes  are quickly depressed
at high energies in MSSM, in agreement with the general expectations  \cite{heli1,heli2}.

On the contrary, the  helicity conserving (HC) amplitudes,  denoted
 as
\bq
 F_{-+-+}~~,~~ F_{+--+}~~,~~ F_{-++-}~~,~~ F_{+-+-}~~, ~~\label{HC-amplitudes}
\eq
  remain appreciable at high energies.
Explicit high energy supersimple expressions for them are given in  Appendix A.
In constructing them, we separate the HC amplitudes into two parts,
defined in  Sect. 2.
The first one, called "augmented Sudakov"
 part, contains  contributions from the  triangles, boxes and
 the electron and top-quark self-energy  counter-terms (c.t.);
 while the second part, called  "augmented renormalization group (RG)" part,  is obtained from
  the $\gamma\gamma$, $\gamma Z$ and $ZZ$ renormalized self-energy bubbles, exchanged in the
$s$-channel.

These two parts  are respectively denoted as
 $F^{\rm Sud}_{\lambda \lambda'\mu\mu'}$ and $F^{\rm s.e.}_{\lambda \lambda'\mu\mu'}$.
As discussed in Sect. 2, the independence of this separation from
 the gauge fixing procedure, is guaranteed by subtracting
 from   $F^{\rm Sud}_{\lambda \lambda'\mu\mu'}$ the pinch part
  of the    triangular  graphs in Fig.\ref{pinch-fig}, and including
  it in $F^{\rm s.e.}_{\lambda \lambda'\mu\mu'}$  \cite{pinch, pinch1}.

In Sect. 3 and  Appendix A,  we discuss our predictions for the
 HV and HC amplitudes for process $e^-e^+\to t\bar t$, in  MSSM models.
 As examples of the way such expressions may be used in studying physically
 observable quantities, we consider
 the differential cross section $d\sigma(e^-e^+\to t\bar t)/d\cos\theta$,
 the forward-backward
 ($A_{FB}$) and the left-right ($A^t_{LR}$) asymmetries.
 It is then argued that the supersimple MSSM expressions for
 the HC amplitudes, may  be useful for quickly distinguishing SUSY contributions from
  possible new physics contributions  induced  e.g. by a new $Z'$ vector or axial boson,
 or by new  anomalous  $Zt\bar t$ couplings.

Finally in the fourth  section we  give our conclusions and discuss the theoretical aspects
and the  implications of our results. \\

\section{The augmented Sudakov and RG forms.}

\begin{figure}[h]
\[
\hspace{-1.cm}
\epsfig{file=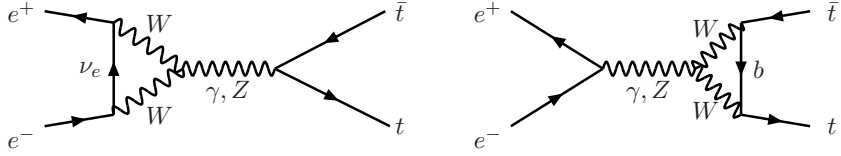, height=2.cm}
\]
\vspace*{-1.cm}
\caption[1]{Diagrams contributing to the pinch term.}
\label{pinch-fig}
\end{figure}

As explained in the Introduction, the  "augmented Sudakov" part
of the HC  amplitudes, denoted as
$F^{\rm Sud}_{\lambda \lambda'\mu\mu'}$, contains the contributions from the
 triangles and boxes, as well as the contributions from the counter-terms (c.t.) related to the
 external $(e^-,e^+,t,\bar t$) particles. To ensure gauge invariance though,
we have subtracted from them, the pinch part of the $W\nu_eW$ and $WbW$ triangles,
  indicated in  Fig.\ref{pinch-fig} \cite{pinch, pinch1}.
 This "pinch" term handling, only affects  $F_{-+-+}$.

Apart from $F^{\rm Sud}_{\lambda \lambda'\mu\mu'}$,
 there exists also the $F^{\rm s.e.}_{\lambda \lambda'\mu\mu'}$
part of the HC amplitudes, called  "augmented  RG" part. This  contains the contributions
from the $\gamma\gamma$, $\gamma Z$ and $ZZ$ renormalized self-energy bubbles in the
$s$-channel, and includes  also  the pinch term mentioned above.

An easy way to calculate both these  amplitude-parts at high energy, is  by studying
the SUSY-transformed
process $ \tilde e^- \tilde e^+\to \tilde t \bar{\tilde t}$ \cite{super}.
But  in order to  unambiguously  obtain
all   constant terms, a direct 1loop computation of the $e^-e^+\to t \bar t $ amplitudes
is also made, following \cite{ttsud} and  using the asymptotic expansion \cite{asPV}
of the Passarino-Veltman (PV)
functions  \cite{PV}.\\

Denoting by  $x,y$, any two of the Mandelstam variables $(s,t,u)$ in $ e^-  e^+\to t \bar  t$,
 while  $V=\gamma,Z,W$,  we find that a supersimplicity renormalization scheme (SRS)
 may  be defined in MSSM. In this SRS scheme,
 the high energy 1loop  HC amplitudes are given by
a linear combination  of the forms
\bqa
&& \overline{\ln^2x_V}\equiv
\ln^2 x_V +2L_{a_1Vc_1}+2L_{a_2Vc_2} ~~, ~~~
x_V \equiv \left (\frac{-x-i\epsilon}{m_V^2} \right ) ~~, \label{Sud-ln2} \\
&& \overline{\ln x_{ij}}\equiv \ln x_{ij}+b^{ij}_0(m_a^2)-2 ~~,~~~
\ln x_{ij}\equiv \ln \frac{-x-i\epsilon}{m_im_j} ~~, \label{Sud-ln} \\
&& \ln^2r_{xy}+ \pi^2 ~~~,~~~  r_{xy} \equiv \frac{-x-i\epsilon}{-y-i\epsilon} ~~~~,
\label{r-form}
\eqa
with  the  coefficients of the Sudakov forms (\ref{Sud-ln2}) and (\ref{Sud-ln})  being
constants, satisfying the  general constraints
\cite{MSSMrules1,MSSMrules2,MSSMrules3,MSSMrules4}; while the
coefficients of (\ref{r-form}) may also contain  ratios of  Mandelstam variables,
as well as   constants. No additional overall terms can exist in the SRS HC amplitudes
\cite{super}.  This structure is exactly the same as in \cite{super}.
Such  SRS HC amplitudes are related
 to  the on-shell renormalization scheme ones  \cite{OS},
 through an additional residual constant contribution  \cite{super}.
 The expressions for the on-shell HC amplitudes thus obtained, are the
 "supersimple" expressions mentioned above and given in Appendix A.\\

We next discuss the forms (\ref{Sud-ln2}-\ref{r-form}).
As shown in \cite{super},  the augmented  Sudakov squared-logs
appearing in (\ref{Sud-ln2}) are  always
associated to triangles or boxes involving  gauge exchanges  $(V=\gamma,Z,W)$.
In particular the  $L_{a_iVc_i}$ terms appearing there,  are defined  by
 \bqa
L_{aVc}\equiv L(p_a, m_V, m_c) & = &
 \phantom{+} {\rm Li_2} \left ( \frac{2p_a^2+i\epsilon}{m_V^2-m_c^2+p_a^2+i\epsilon +
\sqrt{\lambda (p_a^2+i\epsilon, m_V^2, m_c^2)}} \right )
\nonumber \\
&& + {\rm Li_2} \left ( \frac{2p_a^2+i\epsilon }{m_V^2-m_c^2+p_a^2+i\epsilon -
\sqrt{\lambda (p_a^2+i\epsilon, m_V^2, m_c^2)}} \right )~~,
\label{LaVc-term}
\eqa
where ${\rm Li_2}$ is a Spence function and
\bq
\lambda(a,b,c)=a^2+b^2+c^2-2ab-2ac-2bc~~. \label{lambda-function}
\eq
Note  that in  $L_{a_iVc_i}$ in  (\ref{Sud-ln2}), the gauge boson always appears
as a  middle index;   while   $a_i$
describes an external particle of $e^-e^+\to t \bar t $; and $c_i$ denotes an internal exchange
in the diagram generating the specific high energy term
\cite{MSSMrules1,MSSMrules2,MSSMrules3,MSSMrules4}.\\

We next turn to augmented  Sudakov linear logs  in (\ref{Sud-ln}).
The constant contribution $b_0^{ij}(m_a^2)$ in them, is determined by the finite part
of the $B_0^{ij}(m_a^2)$   PV  function \cite{PV, super}
\bqa
b_0^{ij}(m_a^2)& \equiv& b_0(m_a^2; m_i,m_j) =
2 + \frac{1}{m_a^2} \Big [ (m_j^2 -m_i^2)\ln\frac{m_i}{m_j}\nonumber\\
&& + \sqrt{\lambda(m_a^2+i\epsilon, m_i^2, m_j^2)}  {\rm ArcCosh} \Big
(\frac{m_i^2+m_j^2-m_a^2-i\epsilon}{2 m_i m_j} \Big ) \Big ] ~~, \label{b0ij}
\eqa
where   $(i,j)$ describe  two internal exchanges, while $m_a$ denotes
the mass of either an  external particle $(e^\mp, t,\bar t)$,
or a neutral gauge boson $(V=\gamma,Z)$, that can
couple to the $ij$-pair.

The first case, where   $b_0^{ij}(m_a^2)$ is
associated to an external line,   arises from the cancelation between
the divergences
\bq
\Delta-\ln+b_0^{ij}(m_a^2) ~~, \label{linear-ln-Delta}
\eq
induced by  triangular   diagrams, and those induced by
the $e,t$ counter terms (c.t.), finally leading to
expressions like $\ln+b_0^{ij}(m_a^2)-c$,  where $c$ is a pure number.
But then  a remarkable  property appears in SUSY, where only the HC amplitudes need
to be considered \cite{heli1, heli2}. For each group of related diagrams,
the value of $c$ induced by  the SM-exchanges differs from the one induced by the the pure SUSY
ones. It is  only when all related   diagrams are combined,
 that the sum of the SM and  SUSY
contributions produces the value $c=2$  appearing in (\ref{Sud-ln})  \cite{super}.

A few  typical examples are:
\begin{itemize}
\item
For  triangles involving a single gauge exchange  related
 to the initial $e^\mp $ lines and their (c.t.),
the  gauge exchanges contribute  $3\ln +3b_0^{Vf}(m_e^2) -7$, with $(f=e, \nu_e)$ and
$(V=W,Z,\gamma)$. The corresponding  SUSY gaugino-slepton exchanges give
$-\ln -b_0^{\tilde V\tilde f}(m_e^2) +3$. Adding the two,   the MSSM total result becomes
 a combination of  forms like   $[\ln +b_0^{ij}(m_e^2) -2]$.\\

\item
For Yukawa triangles connected to the final $(t, \bar t)$ lines and their (c.t.),
the SM Higgs exchanges produce terms like $-\ln -b_0(m_t^2) +3$, while
the SUSY additional Higgs and higgsino exchanges contribute $-\ln -b_0(m_t^2) +1$, so
 that   the MSSM  total is again a combination of forms like $[\ln +b_0(m_t^2) -2]$.

\end{itemize}

The second case  where  in  (\ref{Sud-ln}) we have  $m_a=m_\gamma,\mz$,
 was never seen in  the  processes studied in \cite{super}.
 It is  first observed  here for $e^-e^+\to t\bar t$.
 Regularizing the infrared singularities
by choosing $m_\gamma=m_Z$   \cite{super}, we thus encounter
additional augmented Sudakov linear logs  like
\bqa
\overline{\ln s_{WW}} &= & \ln s_{WW}+b_0^{WW}(m_z^2)-2 ~~, \nonumber \\
\overline{\ln s_{H^+H^-}} &= & \ln s_{H^+H^-}+b_0^{H^+H^-}(m_z^2)-2 ~~, \nonumber \\
\overline{\ln s_{h^0Z}} &= & \ln s_{h^0Z}+b_0^{h^0Z}(m_z^2)-2 ~~, \nonumber \\
\overline{\ln s_{H^0Z}} &= & \ln s_{H^0Z}+b_0^{H^0Z}(m_z^2)-2 ~~, \nonumber \\
\overline{\ln s_{h^0A^0Z}} &= & \ln s_{h^0A^0}+b_0^{h^0A^0}(m_z^2)-2 ~~, \nonumber \\
\overline{\ln s_{H^0A^0}} &= & \ln s_{H^0A^0}+b_0^{H^0A^0}(m_z^2)-2 ~~,\nonumber \\
\overline{\ln s_{ff}} &= & \ln s_{ff}+b_0^{ff}(m_z^2)-2 ~~,\nonumber \\
\overline{\ln s_{\tilde f_i \tilde f_j}} &= & \ln s_{\tilde f_i \tilde f_j}
+b_0^{\tilde f_i \tilde f_j}(m_z^2)-2 ~~, \nonumber \\
\overline{\ln s_{\tchi_i \tchi_j}} &= & \ln s_{\tchi_i \tchi_j}
+b_0^{\tchi_i \tchi_j}(m_z^2)-2 ~~, \label{b0mZ}
\eqa
where  the  indices $(i,j)$  in $\ln s_{ij}$,  describe particles with non-vanishing
$\gamma ij$ or $Zij$ couplings. Such terms  are  generated by   counter terms
in the $\gamma, Z$ self energy insertions $\Sigma_{\gamma\gamma}(s)$, $\Sigma_{\gamma Z}(s)$
 and $\Sigma_{ZZ}(s)$. More explicitly, the
 gauge self-energy insertions give contributions   like  $-\Delta+\ln-2$,
 whose    $\Delta$-divergence is  canceled   by  quantities like
$\Delta+b^{ij}_0(\mz^2)-\ln (m_im_j/\mu^2)$, induced
by the gauge wave function renormalization constants  \cite{asPV}. This is
similar to the    case discussed around
(\ref{linear-ln-Delta}), where the divergences
are canceled by  the electron  or top  counter terms. \\

We  also remark  that the terms in (\ref{b0mZ})  concern  only the pinch
and the augmented RG parts of the  high energy HC amplitudes.
The augmented  Sudakov contributions to the high energy HC amplitudes,  do not have
this form.

As a result, the $\overline{\ln s_{WW}}$ term  in the
$F^{\rm Sud}_{-+-+}$ expression   (\ref{sim-FSudmpmp}), is directly related
to the subtraction of  the
pinch contribution  from the diagrams in Fig.\ref{pinch-fig} \cite{pinch, pinch1}.
  Its magnitude is given by
\bq
 \frac{\alpha^2}{\sw^4}\, \overline{\ln s_{WW}}  ~~. \label{pinch-term}
 \eq
It is this term that  has been subtracted from the definitions of
 $F^{\rm Sud}_{-+-+}$, and inserted in
 $F^{\rm s.e.}_{-+-+}$ given in  (\ref{sim-Fsempmp}).
None of the other high energy HC amplitudes
$F^{\rm Sud}_{\lambda \lambda'\mu\mu'}$,  is affected by terms in (\ref{b0mZ}).

 In contrast to this,  all forms  (\ref{b0mZ})  contribute
 to the augmented RG parts of the asymptotic HC amplitudes
 $F^{\rm s.e.}_{\lambda \lambda'\mu\mu'}$.\\

Finally, in addition to the augmented  squared and linear logs scaled by masses,
a third form  given by (\ref{r-form}), also appears  in the high energy HC amplitudes
\cite{super}. Typical expressions of this kind, for both
$\tilde e^-\tilde e^+ \to \tilde t \bar{\tilde t}$
and  $e^-e^+\to t \bar t $ processes, are
 $(\ln^2r_{us}+\pi^2)$ or $(\ln^2r_{ts}+\pi^2)$, always arising  purely
 from boxes.\\

\section{The  HC  amplitudes for $e^-e^+\to t \bar t$.}

In this Section  we discuss the exact 1loop EW results for the
$F^{\rm Sud}_{\lambda \lambda'\mu\mu'}$ and $F^{\rm s.e.}_{\lambda \lambda'\mu\mu'}$
parts of the HC amplitudes in MSSM, and compare them  to the corresponding
 high energy supersimple expressions given in Appendices A.1 and A.2 respectively.
The results in (\ref{sim-FSudmpmp}-\ref{sim-FSudpmmp}) and
(\ref{sim-Fsempmp}-\ref{sim-Fsepmmp}) clearly indicate that the Yukawa interactions and the RG
contributions, do respect the supersimplicity structure.

As we show below,  these supersimple expressions  reproduce
   the main features  of the exact 1loop amplitudes,
   even for  energies close to the SUSY scale.

 For assessing this explicitly, we first show
 the quick vanishing, as the energy increases,  of the HV amplitudes.
 Then, we turn to the augmented Sudalov part of the  HC amplitudes and
 compare the exact 1loop results for
 $F^{\rm Sud}_{\lambda \lambda'\mu\mu'}$, with the corresponding supersimple high
 energy expressions in Appendix A.1. And once this is done, we turn to the complete amplitudes
\bq
F_{\lambda \lambda'\mu\mu'}=F^{\rm Born}_{\lambda \lambda'\mu\mu'}
+F^{\rm Sud}_{\lambda \lambda'\mu\mu'}+
F^{\rm s.e.}_{\lambda \lambda'\mu\mu'}~~, \label{HC-complete}
\eq
 and compare them to their supersimple approximation obtained by summing
 the corresponding expressions  in Appendix A.\\

For the numerical  illustrations,  we use two MSSM benchmarks, consistent with the present
LHC results. The first, called MSSMhigh, is given by the  cMSSM high scale
parameters \cite{Buch}
\bq
m_0=1080 ~~,~~m_{1/2}=1800~~,~~ A_0=860~~,~~ \tan\beta=48~~,~~ \mu>0 ~~, \label{model-MSSMhigh}
\eq
where all dimensional quantities are  in GeV. For this model, the  SuSpect code  gives
 $m_{h^0}\simeq 122 ~{\rm GeV}$, while
the lightest neutralino is put at about 800 GeV, and  all other SUSY particles acquire
masses between 1000 and almost 4000 GeV \cite{suspect}. As a result, the SUSY contribution
to $(g_\mu-2)/2$ is tiny, in this benchmark. \\

The second benchmark, called MSSMlow, is characterized be the EW scale
parameters \cite{MSSMlow1, MSSMlow2}
\bqa
&& M_1=100~~,~~ M_2=200~~,~~ M_3=800 ~~, \nonumber \\
&& m_{\tilde l} =400 ~~,~~ m_{\tilde q}=1100~~,~~
 A_{\tau} =-800 ~~,~~ A_b=A_t=- 2200 ~~,~~ \nonumber \\
&& \mu=200 ~~,~~ m_{A^0}=320~~,~~ \tan\beta =20~~ , \label{model-MSSMlow}
\eqa
where $m_{\tilde l}, m_{\tilde q}$ describe the common EW scale
SUSY breaking slepton and  squark masses, for all three generations; (again all masses in GeV).
 The charginos, neutralinos and sleptons in  MSSMlow  are much lighter than in  the
 previous benchmark. Consequently, this benchmark can accommodate a large SUSY contribution
 to $(g_\mu-2)/2$, consistent with the experimental data \cite{amu1, amu2}.
Moreover, SuSpect  \cite{suspect} gives for it $m_{h^0}\simeq 125~ {\rm GeV}$, the lightest
neutralino is put at 90 GeV, and  the $m_{A^0}$ and $m_{H^0}$ masses are
 in the 320 GeV region  \cite{MSSMlow2}.\\

Using these two MSSM benchmarks\footnote{A very short list of other possible
 benchmarks may be found
in \cite{otherLHC}.}, we present in  Fig.\ref{HV-fig},
the two independent HV amplitudes $F_{-+--},~F_{+---}$, as functions of energy,
at a fixed c.m. angle $\theta=60^o$.  As seen there, the 1loop EW order results
for both HV amplitudes, as well as their Born approximation,
are almost identical and quickly suppressed at high energies,
in agreement with the general helicity-conservation
(HCns)  theorem \cite{heli1,heli2}.

We conclude therefore that  for a quick study of physical observables,
it may be sufficient to use the Born approximation for the HV amplitudes. \\

\subsection{The $F^{\rm Sud}_{\lambda \lambda'\mu\mu'}$ part of the HC amplitudes.}

We first investigate whether the exact 1loop results for $F^{\rm Sud}_{\lambda \lambda'\mu\mu'}$
agree  with the corresponding supersimple expressions, at asymptotic energies.
In other words, whether there are any residual contributions that they should still be added
to the expressions in Appendix A.1.  Such residual contributions are
 essentially determined by the differences between  the  $e$ or $t$
 wave function renormalization constants,  in the on-shell \cite{OS}
and SRS renormalization schemes \cite{super},
\bq
\delta Z^{L,R,~{\rm res}}_f=Z^{L,R,~{\rm OS}}_f-Z^{L,R,~{\rm SRS}}_f ~~,\label{deltaZf-def}
\eq
where $f=e,~t$. For this we find\\
for MSSMlow  (\ref{model-MSSMlow})
\bqa
\delta Z^{L,~{\rm res}}_e=-0.00091 ~~ &,& ~~\delta Z^{R,~{\rm res}}_e=-0.00243~~, \nonumber \\
\delta Z^{L,~{\rm res}}_t=0.00202 ~~ &,& ~~ \delta Z^{R,~{\rm res}}_t=0.00196
~~, \label{deltaZf-low}
\eqa
while for MSSMhigh  (\ref{model-MSSMhigh})
\bqa
\delta Z^{L,~{\rm res}}_e=-0.00039 ~~ &,& ~~\delta Z^{R,~{\rm res}}_e=-0.00124~~, \nonumber \\
\delta Z^{L,~{\rm res}}_t=0.00330 ~~ &,& ~~ \delta Z^{R,~{\rm res}}_t=0.00051
~~.  \label{deltaZf-high}
\eqa
Thus $|\delta Z^{L,R,~{\rm res}}_f|\ll 1$, which means that no further residual terms
are needed  in (\ref{sim-FSudmpmp}-\ref{sim-FSudpmmp}).\\

In Fig.\ref{Sud-fig}  we then present the energy dependence
of the  augmented Sudakov part of the HC amplitudes
 $F^{\rm Sud}_{\lambda \lambda'\mu\mu'}$. The c.m. scattering angle is fixed at $\theta=60^o$.
Full lines describe the  exact 1loop EW order results;
while broken lines,  indicated by "sim", denote the supersimple high energy
 amplitudes  in Appendix A.1.

As seen in this figure, the differences between the exact
and  supersimple results, are almost invisible for all HC augmented Sudakov
amplitudes, at all energies, for the MSSM models (\ref{model-MSSMlow}, \ref{model-MSSMhigh}).
In fact, at energies in the range
$ 0.4 \lesssim \sqrt{s} \lesssim  1 $ TeV, some visible differences only appear for
$F^{\rm Sud}_{-+-+}$; but they become invisible for $\sqrt{s} \gtrsim 1$ TeV.

Therefore, the  the supersimple expressions
 for  the  augmented  Sudakov amplitudes in Appendix A.1, approach the corresponding
 exact 1loop results, very quickly, for the above MSSM benchmarks. \\

\subsection{The $F^{\rm s.e.}_{\lambda \lambda'\mu\mu'}$ part of the HC amplitudes. }

As already said in Sect.2, the  augmented RG part for the HC amplitudes
$F^{\rm s.e.}_{\lambda \lambda'\mu\mu'}$, describes  the 1loop finite contribution
arising from the  renormalized $\gamma\gamma$, $\gamma Z$ and $ZZ$
self-energy functions, together with  the pinch contribution of the graphs in Fig.\ref{pinch-fig}.
The high energy supersimple expressions for these $e^- e^+\to t \bar t $
amplitudes appear in Appendix A.2.

Using the definitions in  Appendix A.2 and (\ref{F-Born-HC}),
we first check that the logarithms  of this  part
coincide with those in the renormalization group  result
\bq
F^{\rm RG~log}=- {1\over 4\pi^2} \ln \left ({s\over m^2_Z}\right )
\left [\beta_2g^4_2\left ({dF^{Born}\over dg^2_2}\right )
+\beta_1g^4_1 \left ({dF^{Born}\over dg^2_1} \right ) \right ] ~~, \label{F-RG1}
\eq
where $g_1=e/c_W$, $g_2=e/s_W$ and
\bqa
 && \beta_1={-11\over4}~~~~,~~~~\beta_2={-1\over4}~~~~, \label{beta-RG}
\eqa
 leading to
\bqa
F^{\rm RG~log}_{-+-+} &= & \alpha^2 \left ({2u\over s}\right )
\left [ {-3+6s^2_W-14s^2_W\over12s^4_Wc^4_W}\right ]
\ln \left ({s\over m^2_Z} \right ) ~~, \nonumber \\
F^{\rm RG~log}_{+-+-} &= & \alpha^2 \left ({2u\over s} \right )
\left [ {-22\over3c^4_W} \right ] \ln \left ({s\over m^2_Z} \right ) ~~, \nonumber \\
F^{\rm RG~log}_{-++-} &= & \alpha^2 \left ({-2t\over s} \right )
\left [{11\over3c^4_W}\right ] \ln \left ({s\over m^2_Z} \right )  ~~, \nonumber \\
F^{\rm RG~log}_{+--+} &= & \alpha^2 \left ({-2t\over s} \right )
\left [{11\over6c^4_W} \right ] \ln \left ({s\over m^2_Z} \right ) ~~ , \label{F-RG2}
\eqa
which indeed agree with the logarithmic terms in (\ref{sim-Fsempmp}-\ref{sim-Fsepmmp}).\\

We next discuss the energy independent  residual terms, that are needed
in the supersimple expressions
(\ref{sim-Sigma-gamgam},\ref{sim-Sigma-gamZ},\ref{sim-Sigma-ZZ}); in order to
describe the exact 1loop values for
  $\hat \Sigma_{\gamma\gamma}$, $\hat \Sigma_{\gamma Z}$, $\hat \Sigma_{ZZ}$,
at asymptotic energies.

For $\hat{\Sigma}_{\gamma\gamma}(s)$, no  such  term is needed  in
(\ref{sim-Sigma-gamgam}).

But for  $\hat{\Sigma}_{\gamma Z}(s)$ and $\hat{\Sigma}_{ZZ}(s)$, a quantity like
\bq
\overline{\Delta\rho} \simeq 0.017 ~~, \label{Delta-rho-bar}
\eq
is needed in (\ref{sim-Sigma-gamZ},\ref{sim-Sigma-ZZ}), for the  MSSM benchmarks
  (\ref{model-MSSMlow}, \ref{model-MSSMhigh});  similar results at the percent level, are also
  true for the benchmarks in \cite{otherLHC}.   This value is
close to the well-known neutral-to-charged current ratio parameter
\bq
\Delta\rho={\Sigma^{ZZ}(0)\over m^2_Z}-{\Sigma^{WW}(0)\over m^2_W}\sim 0.01 ~~,
\label{Delta-rho}
\eq
mainly determined by  the $(b,t)$ contributions.
Such a similarity is  not  accidental, since  the structure of
(\ref{hatSigmaVV}, \ref{gauge-ct})
suggests that gauge self-energy differences
like those in (\ref{Delta-rho}), play an important role in determining the value of
 $\overline{\Delta\rho}$,  thereby  motivating its name.\\

Taking into account the  $\overline{\Delta\rho}$-contributions
 in  (\ref{sim-Sigma-gamZ},\ref{sim-Sigma-ZZ}),
the differences
between the exact 1loop contribution to the $F^{\rm s.e.}_{\lambda \lambda'\mu\mu'}$
part of the HC amplitudes, and  the supersimple  expressions
(\ref{sim-Fsempmp}-\ref{sim-Fsepmmp}), normalized to the corresponding
Born contributions,  are
\begin{itemize}
\item
for  MSSMlow  (\ref{model-MSSMlow})
\bqa
\delta F^{\rm s.e.}_{-+-+}/F^{\rm Born}_{-+-+}=0.00054~~~ &,&~~~
\delta F^{\rm s.e.}_{+-+-}/F^{\rm Born}_{+-+-}=-0.00303~~, \nonumber  \\
\delta F^{\rm s.e.}_{-++-}/F^{\rm Born}_{-++-}=0.00266~~~&, &~~~
\delta F^{\rm s.e.}_{+--+}/F^{\rm Born}_{+--+}=0.01406~~ , \label{residual-MSSMlow}
\eqa
\item
while for MSSMhigh  (\ref{model-MSSMhigh})
\bqa
\delta F^{\rm s.e.}_{-+-+}/F^{\rm Born}_{-+-+}=-0.00128~~~&,& ~~~
\delta F^{\rm s.e.}_{+-+-}/F^{\rm Born}_{+-+-}=-0.00413 ~~, \nonumber  \\
\delta F^{\rm s.e.}_{-++-}/F^{\rm Born}_{-++-}= 0.00162 ~~~&,& ~~~
\delta F^{\rm s.e.}_{+--+}/F^{\rm Born}_{+--+}=0.01313~~ . \label{residual-MSSMhigh}
\eqa
\end{itemize}
\vspace{0.5cm}

The results  (\ref{residual-MSSMlow}, \ref{residual-MSSMhigh})
 guarantee   that the supersimple expressions (\ref{sim-Fsempmp}-\ref{sim-Fsepmmp}) accurately
 approximate the exact 1loop results  for the $F^{\rm s.e.}_{\lambda \lambda'\mu\mu'}$ HC
 amplitudes at high energies. That is,  no further residual terms  are needed  in
 (\ref{sim-Fsempmp}-\ref{sim-Fsepmmp}), at least   for the two above benchmarks.\\

\subsection{The complete HC  amplitudes}

We next turn to the complete HC amplitude given in (\ref{HC-complete}).

In  Fig.\ref{HC-amp-fig}, we present the energy dependence at $\theta=60^o$,
of the complete HC amplitudes
$F_{-+-+},~F_{+-+-}$ (upper panels), and $F_{-++-},~F_{+--+}$ (lower panels),
in the  benchmarks MSSMhigh (\ref{model-MSSMhigh}) and MSSMlow (\ref{model-MSSMlow}).
For comparison, the exact 1loop SM results are also given.
 Left panels show the 1loop effects on Born, in SM and MSSM; note
 that above 1 TeV, the 1loop effects strongly depend on
 the HC amplitude considered, acquiring their largest values for $F_{-+-+}$.
Right panels  give a feeling of how accurately the supersimple expressions
 approximate the exact 1loop results
 in the energy range  from  the  $t\bar t$-threshold to  7 TeV,
  for the aforementioned MSSM benchmarks.
For $F_{+-+-}, ~F_{+--+}$ this  accuracy is   rather good for both benchmarks.
For MSSMlow, good accuracy also exists for both $F_{-+-+},~F_{-++-}$.
For MSSMhigh though,  discrepancies at the 1\% level persist for $F_{-+-+}$, even for
$\sqrt{s}\gtrsim 7 \,{\rm TeV}$;  while for $F_{-++-}$, the accuracy improves at
$\sqrt{s}\gtrsim 4.5 \, {\rm TeV}$. These  features are due to the high  value (around 3 TeV)
 of the SUSY scale in MSSMhigh, which delays  the vanishing  of the mass-suppressed
contributions. \\

In Fig.\ref{sigma-e-fig}, we give illustrations for the energy
 dependence of the "dimensionless cross section" defined as
\bq
\sum_{\lambda \lambda'\mu \mu'} |F_{\lambda \lambda'\mu\mu'}(e^-e^+\to t\bar t)|^2
 ~~. \label{sig-reduced}
\eq
Full lines give the exact 1loop EW order results, while the broken lines
give the "sim" predictions. By "sim" in the  case of (\ref{sig-reduced})  we mean that,
 the the supersimple results of Appendix A are used for the  HC amplitudes,
 while for  the HV ones the  Born expressions are used.

As seen in the left panel of Fig.\ref{sigma-e-fig},
  the exact and "sim" contributions  for MSSMlow, are very similar.
For MSSMhigh though, the right panel of Fig.\ref{sigma-e-fig}
indicates  a change of sign for the (exact-"sim") difference at  around 3 TeV,
again related to the value of the SUSY scale in this benchmark;
compare the  right panels Fig.\ref{HC-amp-fig}.

Similar patterns for MSSMlow and MSSMhigh also appear in Fig.\ref{sigma-a-fig} and
Fig.\ref{sigma-dz-fig}, where the  angular dependence of the
"dimensionless cross section" (\ref{sig-reduced}) are shown. For MSSMhigh in particular,
the (exact-"sim") difference is  at the 2\% level for 1 TeV c.m. energy,
while it reaches the 1\% level at about 10 TeV.\\

In order to show how these supersimple expressions can be used for quickly disentangling the
supersymmetric effects, from other possible non standard contributions, we now consider
two such examples: an anomalous $Zt\bar t$ coupling described by the effective
interaction in (\ref{dZ}); and an additional $Z'$ with purely vector or axial couplings
to electrons and top quarks (\ref{Zprime}). Such a possibility of anomalous top
properties is open, after the Tevatron recent results \cite{AFBttCDF, AFBttD0}.\\

In Fig.\ref{sigma-dz-fig}, we give the results for the case of
an anomalous $Zt\bar t$ coupling (\ref{dZ}) with $d^Z=\pm 0.15$,
causing the   $\sin\theta$-proportional contribution to the HV amplitudes given in (\ref{F-dZ}).
As seen there, such a $d^Z$  induces  discrepancies, which are
much larger and have a different structure
from those of the (exact-"sim") differences caused by  MSSMlow or MSSMhigh, alone.
Thus,  the supersimple expressions may be adequate for excluding such anomalous couplings.\\

\begin{table}[h]
\begin{center}
{ Table 1: The  $A_{FB}$ and $A^t_{LR}$ asymmetries for $e^-e^+\to t \bar t$,
in two MSSM benchmarks, at the exact 1loop EW order and the "sim" approximation.
The results for adding
to the exact 1loop predictions, a new physics contribution, are also shown.}\\
 \begin{small}
\begin{tabular}{||c|c|c|c|c|c|c||}
\hline \hline
\multicolumn{7} {||c||}{$A_{FB}$}\\
\hline
  & 1loop & SIM  & $d^z=0.15$  & $d^z=-0.15$ & $Z'(V)$ & $Z'(A)$   \\ \hline
${\rm MSSM_{high}}$ \cite{Buch} & 0.855 & 0.859 & 0.776 & 0.725 & 0.813& 0.916 \\
${\rm MSSM_{low}}$ \cite{MSSMlow1, MSSMlow2} & 0.868 & 0.859 & 0.790 & 0.743 & 0.828& 0.921 \\
  \hline \hline
 \multicolumn{7} {||c||}{$A^t_{LR}$}\\
\hline
  & 1loop & SIM  & $d^z=0.15$  & $d^z=-0.15$ & $Z'(V)$ & $Z'(A)$   \\ \hline
${\rm MSSM_{high}}$ \cite{Buch} & 0.271 & 0.293 & 0.237 & 0.219 & 0.222& 0.279 \\
${\rm MSSM_{low}}$ \cite{MSSMlow1,MSSMlow2} & 0.264 & 0.287 & 0.232 & 0.216 & 0.218& 0.266 \\
  \hline \hline
\end{tabular}
 \end{small}
\end{center}
\end{table}

Such a  $\sin\theta$-proportional contribution to the HV amplitudes,
as in  (\ref{dZ},\ref{F-dZ}),
when combined with the MSSM contributions, may also change  the forward-backward
asymmetry $A_{FB}$, to which we now turn.

In addition to   $A_{FB}$, we also  consider the $A^t_{LR}$ Left-Right asymmetry
defined as
\bq
A^t_{LR}\equiv \frac{\sigma(e^-e^+\to t_L \bar t)-\sigma(e^-e^+\to t_R \bar t)}
{\sigma(e^-e^+\to t_L \bar t)+\sigma(e^-e^+\to t_R \bar t)}~~,\label{AtLR}
\eq
where $\sigma(e^-e^+\to t_L \bar t)$  and $\sigma(e^-e^+\to t_R \bar t)$
describe the cross sections for the production
of a $t$-quark with helicities  $\mu=-1/2$ and $\mu=+1/2 $, respectively.
All other polarizations  in  (\ref{AtLR})  are summed over.

The results for $A_{FB}$ and $A^t_{LR}$ are presented
in  Table 1. In detail: the second column gives the exact EW
1loop results for  MSSMhigh and MSSMlow; the third column
gives the corresponding "sim" results, defined as in
the Figs.\ref{sigma-a-fig},\ref{sigma-dz-fig};
the fourth and fifth columns give the effects of adding to the exact 1loop results,
the anomalous HV amplitudes (\ref{F-dZ}), with $d^Z=\pm 0.15$; the sixth column gives
the corresponding effect in  case
the only additional  physics, beyond MSSM, consists of a $Z'$  at 3TeV, coupled  like
in (\ref{Zprime}),  with identical vector couplings to both $e^-e^+$ and $t\bar t$;
while finally the seventh column gives the corresponding effect for an axial $Z'$.

Table 1 reaffirms the implications from
Figs.\ref{sigma-e-fig},\ref{sigma-a-fig},\ref{sigma-dz-fig}.
The  "sim" results,  approximate
the exact 1loop ones, for MSSMlow or MSSMhigh, sufficiently well;
so that a $d^Z=\pm 0.15$ can be distinctly visible, even when the SUSY
implications are  described just by "sim".

Table 1 suggests that this is also true for discovering
a 3 TeV $Z'$ vector or axial contribution, of the kind appearing in (\ref{Zprime}).

The above two examples were chosen with arbitrary values of their parameters,
just in order to illustrate the possibility to use the supersimple expressions, for
detecting  types of physics beyond MSSM.

\section{Conclusions}

In this paper, we have extended the supersimplicity concept
to the MSSM process   $e^-e^+\to t \bar t $,
where    Yukawa interactions and renormalization group (RG)
contributions play important roles. Such features  do not exist in
the originally considered processes in \cite{super}.

More explicitly, the "augmented Sudakov" structure found in \cite{super},
is also observed for the Yukawa part of the electroweak corrections. And the
"augmented RG" structure induced by the  photon and $Z$ exchanges in the $s$-channel,
together with the related pinch contributions, are also found to respect this
supersimplicity structure, with specific $\Delta\rho$ type residual contributions.

This supersimplicity realization is due to spectacular SUSY properties,
arising from cancelations
of virtual standard and spartner contributions,  allowing to write simple
expressions for the helicity conserving amplitudes at high energies.
We have thus obtained  very simple   expressions expressions
for the  high energy on-shell HC amplitudes,
which we have termed "supersimple". At such high energies, the helicity violating
amplitudes are found to be very small. \\

A numerical comparison of the "supersimple" expressions,
  with the exact  1loop results,  shows that their accuracy
is very good,  even  at energies comparable to  the SUSY scale;
at least for the two  benchmarks MSSMhigh and MSSMlow, we have used in the  illustrations.
Both, the energy dependence and the angular distribution of the
cross section presented respectively in Fig.\ref{sigma-e-fig} and
Figs.\ref{sigma-a-fig},\ref{sigma-dz-fig},
 are very well reproduced. Only close to threshold,  one may observe some (small) departures.
 This should remain  true for any  MSSM benchmark, provided the energy is sufficiently
 above the SUSY  scale.

\begin{figure}[u]
\[
\hspace{-1.cm}
\epsfig{file=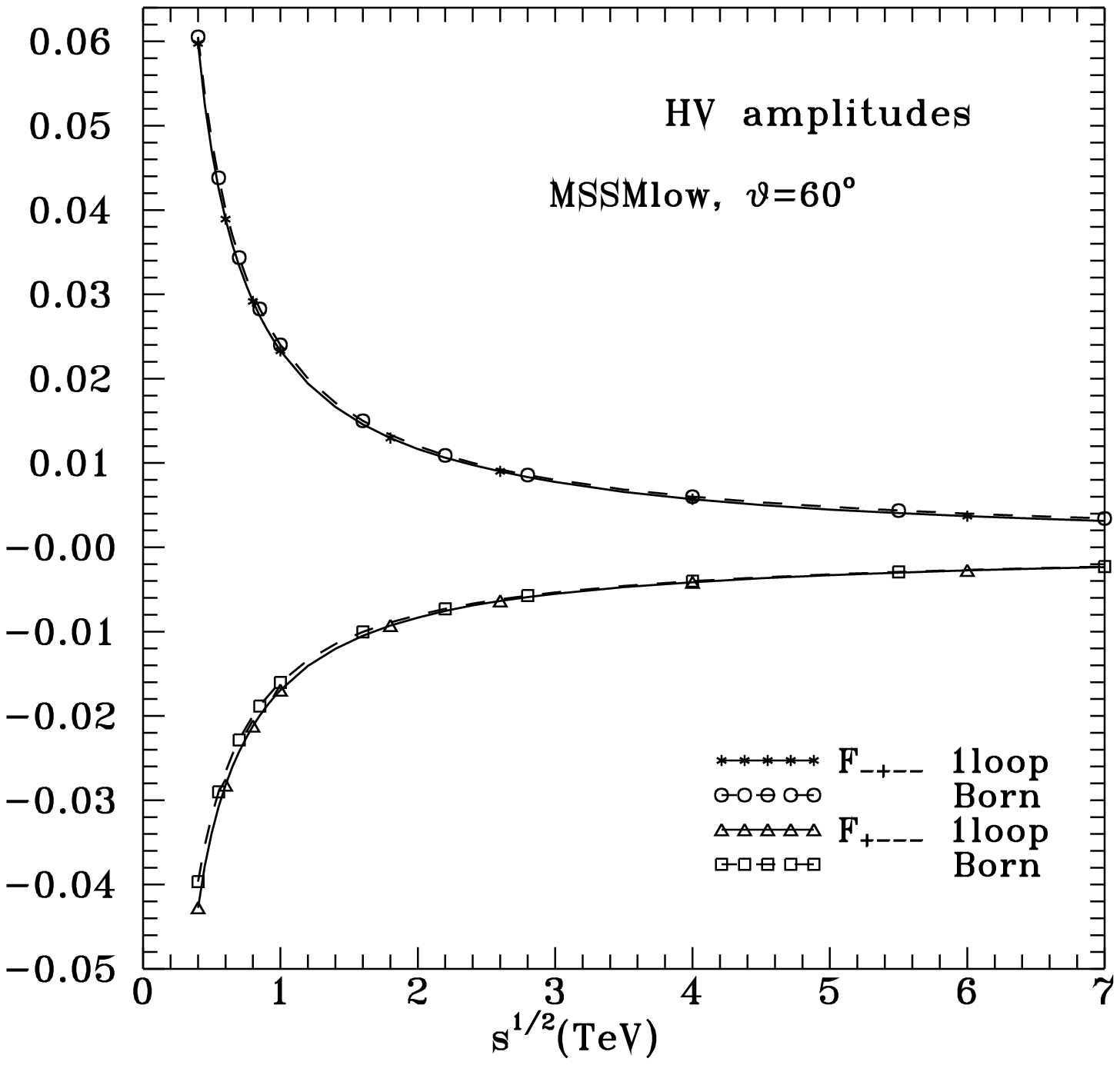, height=7.cm}\hspace{1.cm}
\epsfig{file=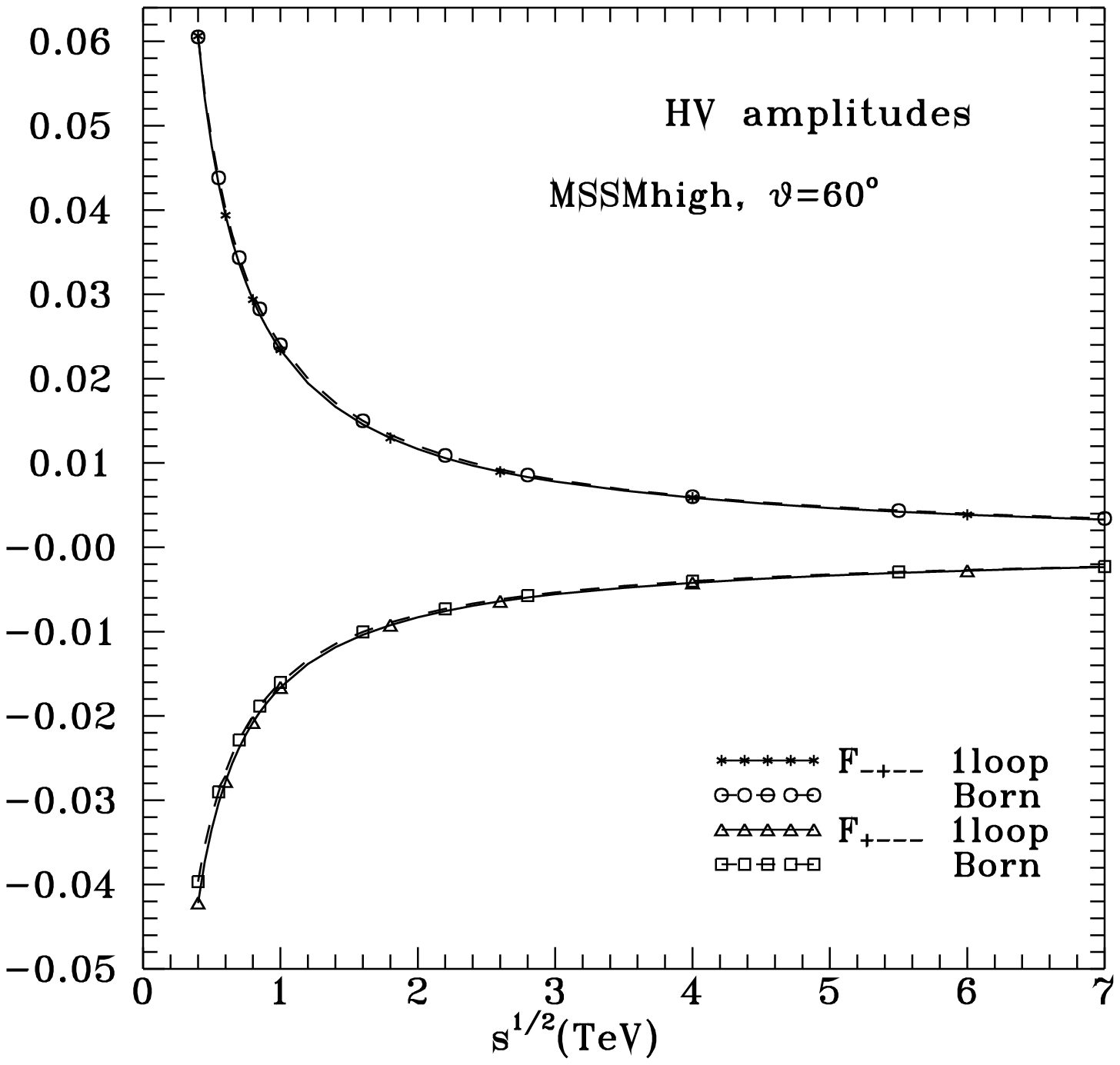,height=7.cm}
\vspace*{-0.5cm}
\]
\caption[1]{Energy dependence at $\theta=60^o$, of the HV amplitudes  $F_{-+--}$ and $F_{+---}$,
at  the 1loop EW order and
their Born approximation. Left panel corresponds to
MSSMlow, defined in (\ref{model-MSSMlow}),  and  right  panel corresponds to
 MSSMhigh, defined  in (\ref{model-MSSMhigh}).}
\label{HV-fig}
\end{figure}

\begin{figure}[d]
\vspace*{-0.5cm}
\[
\hspace{-1.cm}
\epsfig{file=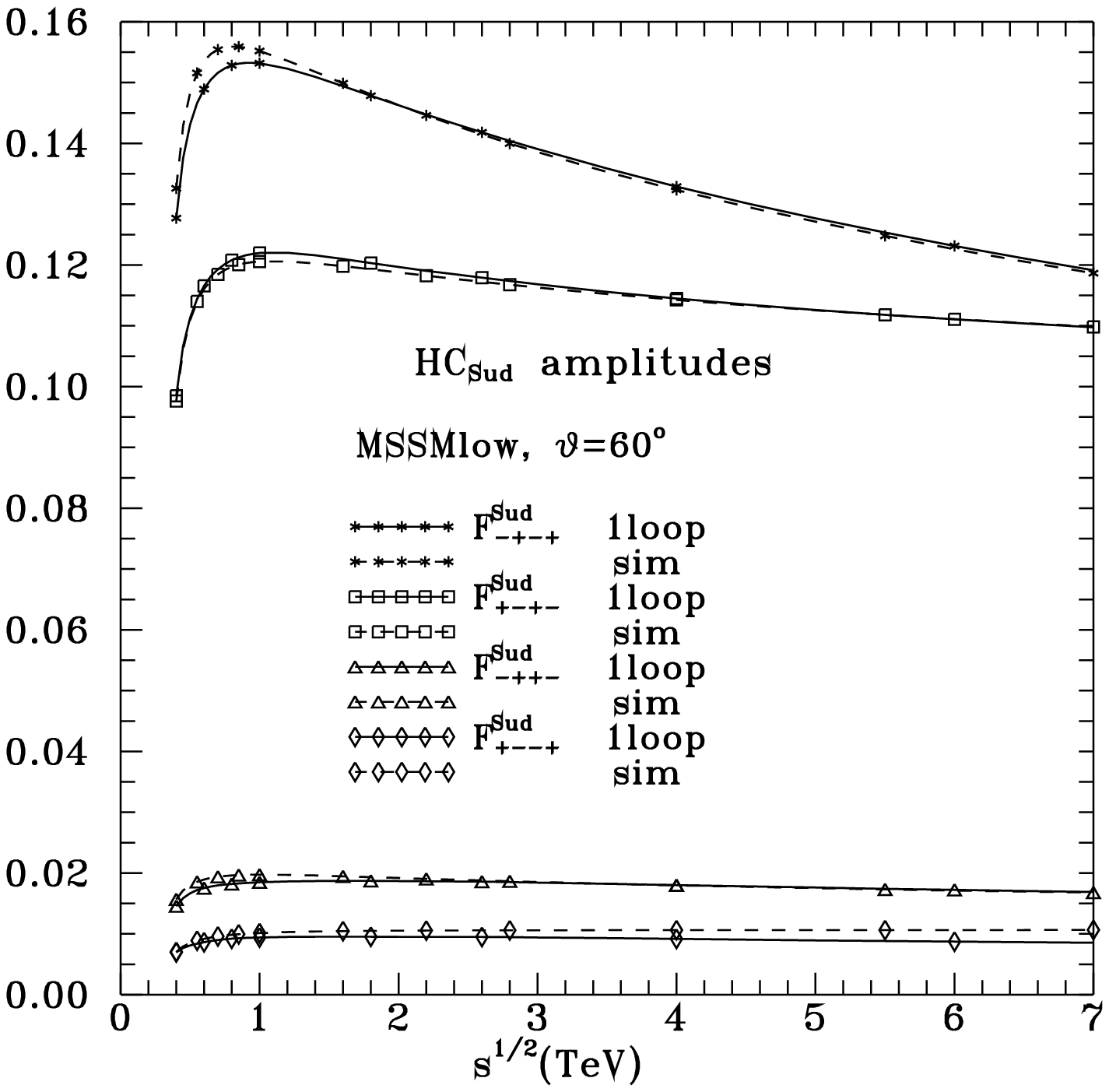, height=7.cm}\hspace{1.cm}
\epsfig{file=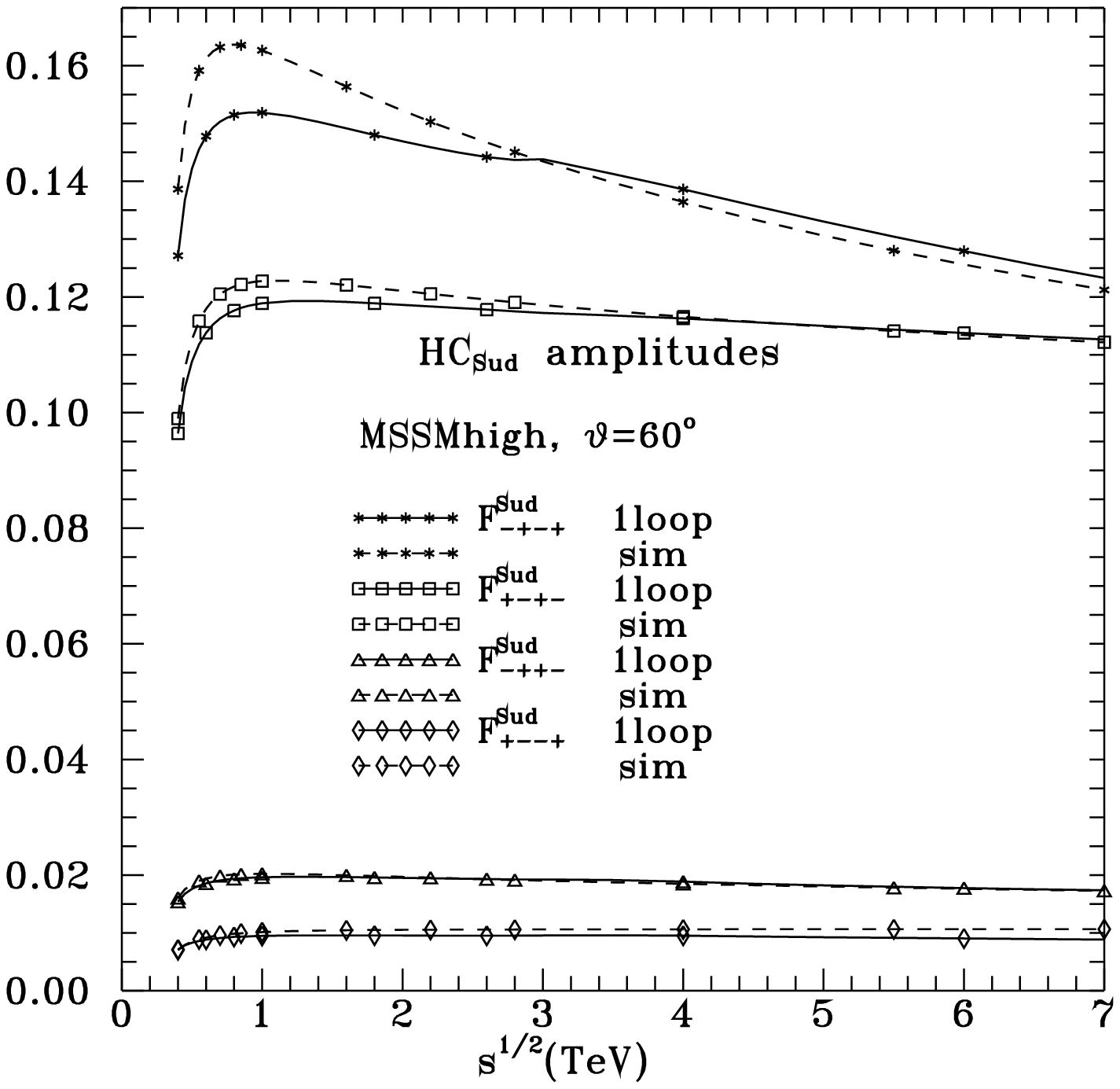,height=7.cm}
\vspace*{-0.5cm}
\]
\caption[1]{Energy dependence at $\theta=60^o$, of the augmented Sudakov part of the
 HC amplitudes  $F^{\rm Sud}_{-+-+}$, $F^{\rm Sud}_{+-+-}$,
 $F^{\rm Sud}_{-++-}$, $F^{\rm Sud}_{+--+}$.
Full lines describe the  exact 1loop EW order results; while broken lines, indicated by "sim",
 denote   the supersimple high energy  approximation given in Appendix A.1.
 Panels and  models  as in  Fig.\ref{HV-fig}.}
\label{Sud-fig}
\end{figure}

\begin{figure}[p]
\vspace*{-0.5cm}
\[
\hspace{-1.cm}
\epsfig{file=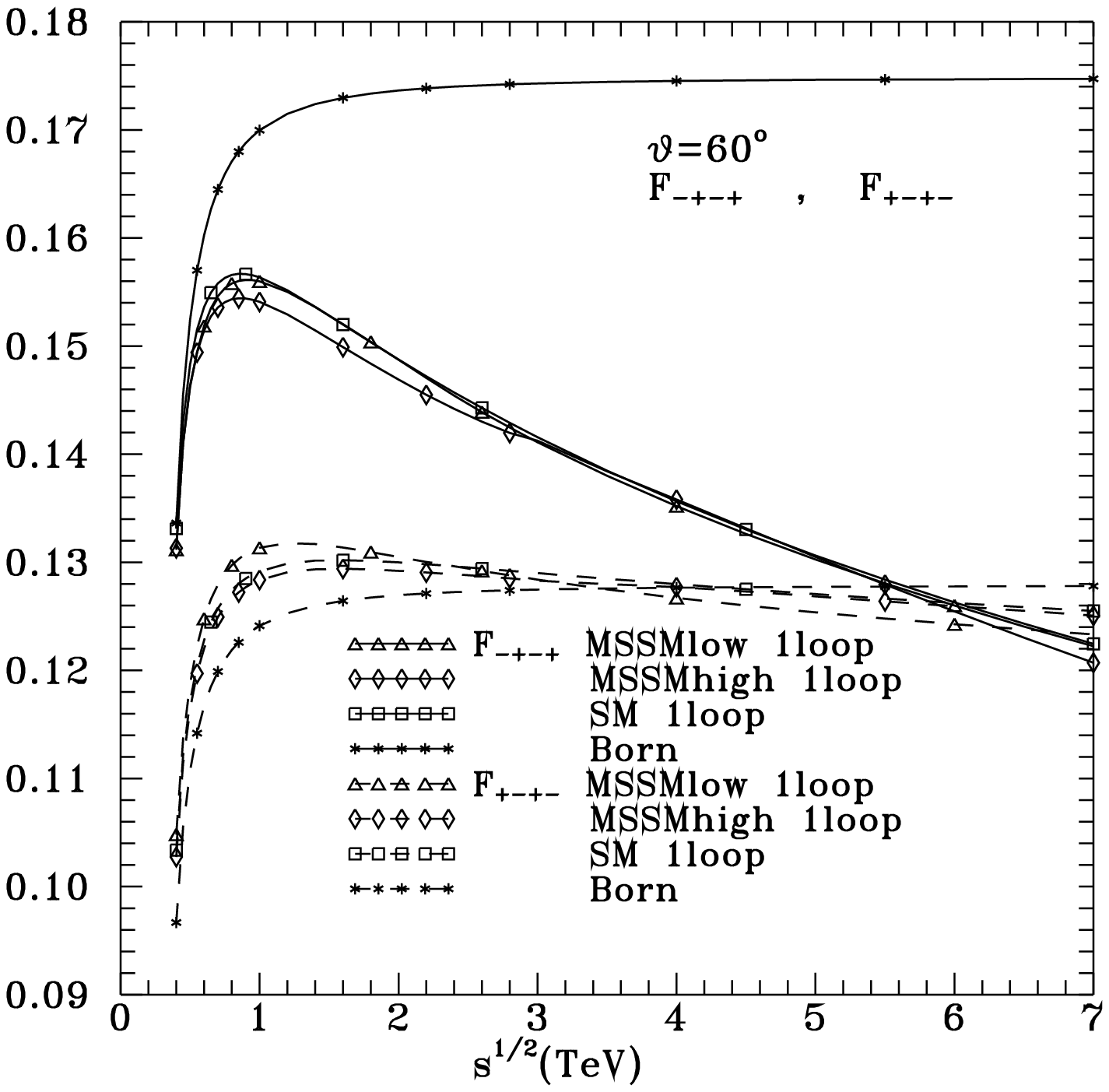, height=7.cm}\hspace{1.cm}
\epsfig{file=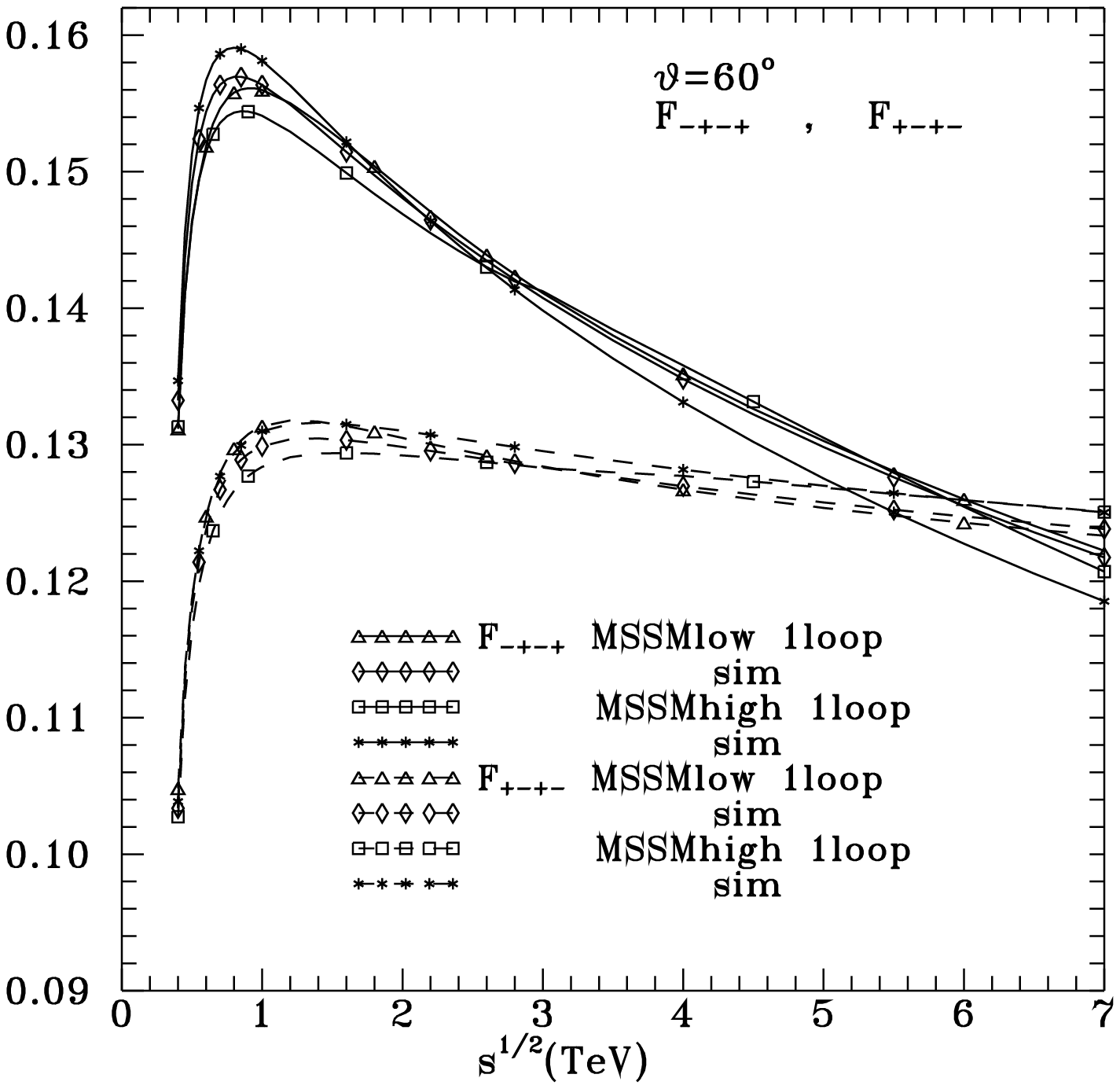,height=7.cm}
\]
\vspace*{0.1cm}
\[
\hspace{-1.cm}
\epsfig{file=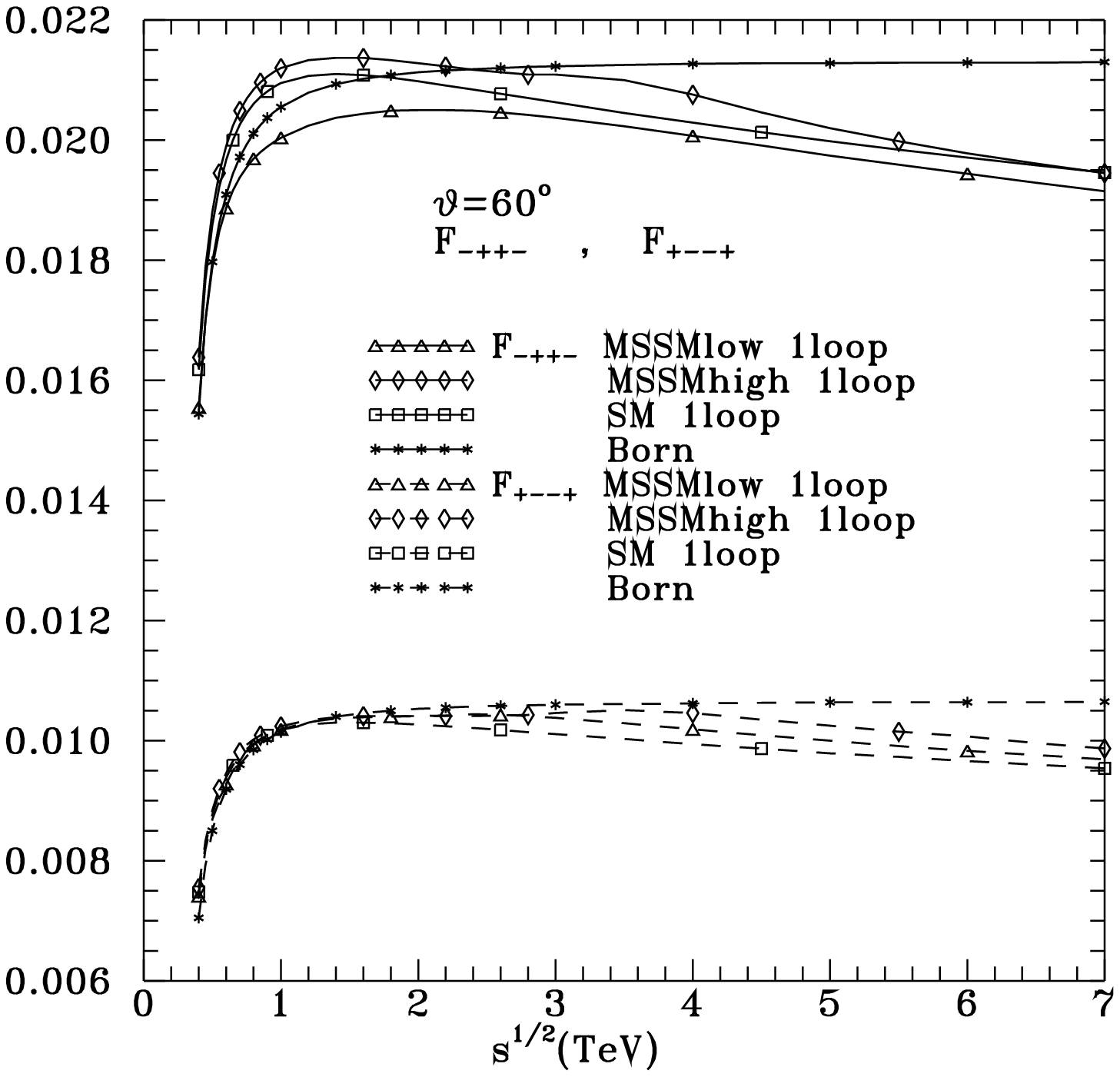, height=7.cm}\hspace{1.cm}
\epsfig{file=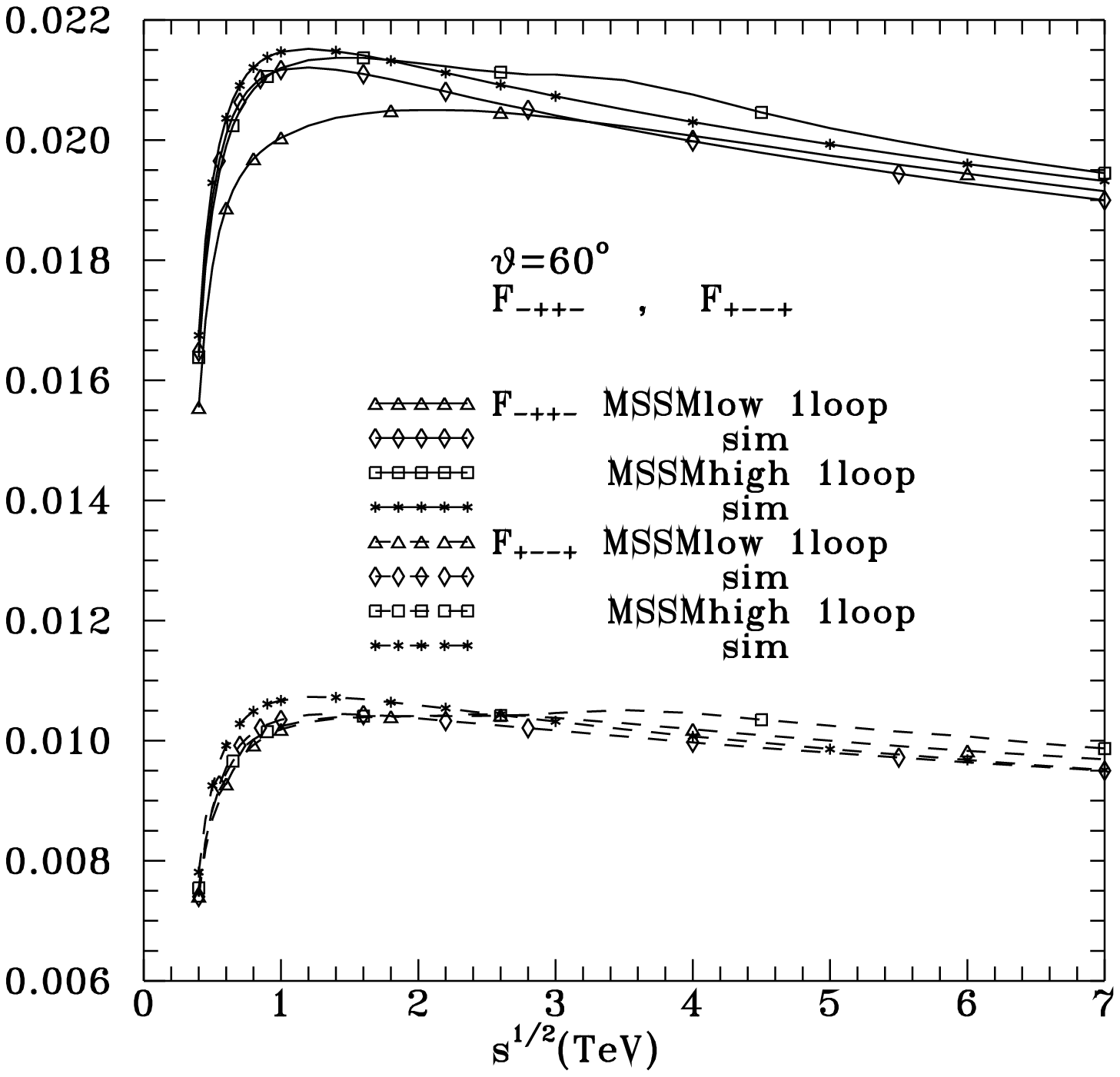, height=7.cm}
\]
\caption[1]{Energy dependence at $\theta=60^o$, of the complete HC amplitudes
$F_{-+-+},~F_{+-+-}$ (upper panels), and $F_{-++-},~F_{+--+}$(lower panels). Models
as in caption of  Fig.\ref{HV-fig}. Left panels show the exact
1loop effects on Born, in SM and MSSM.
Right panels show the accuracy of the supersimple expressions
of Appendix A, indicated  by "sim".}
\label{HC-amp-fig}
\end{figure}

\begin{figure}[u]
\[
\vspace*{-0.5cm}
\hspace{-1.cm}
\epsfig{file=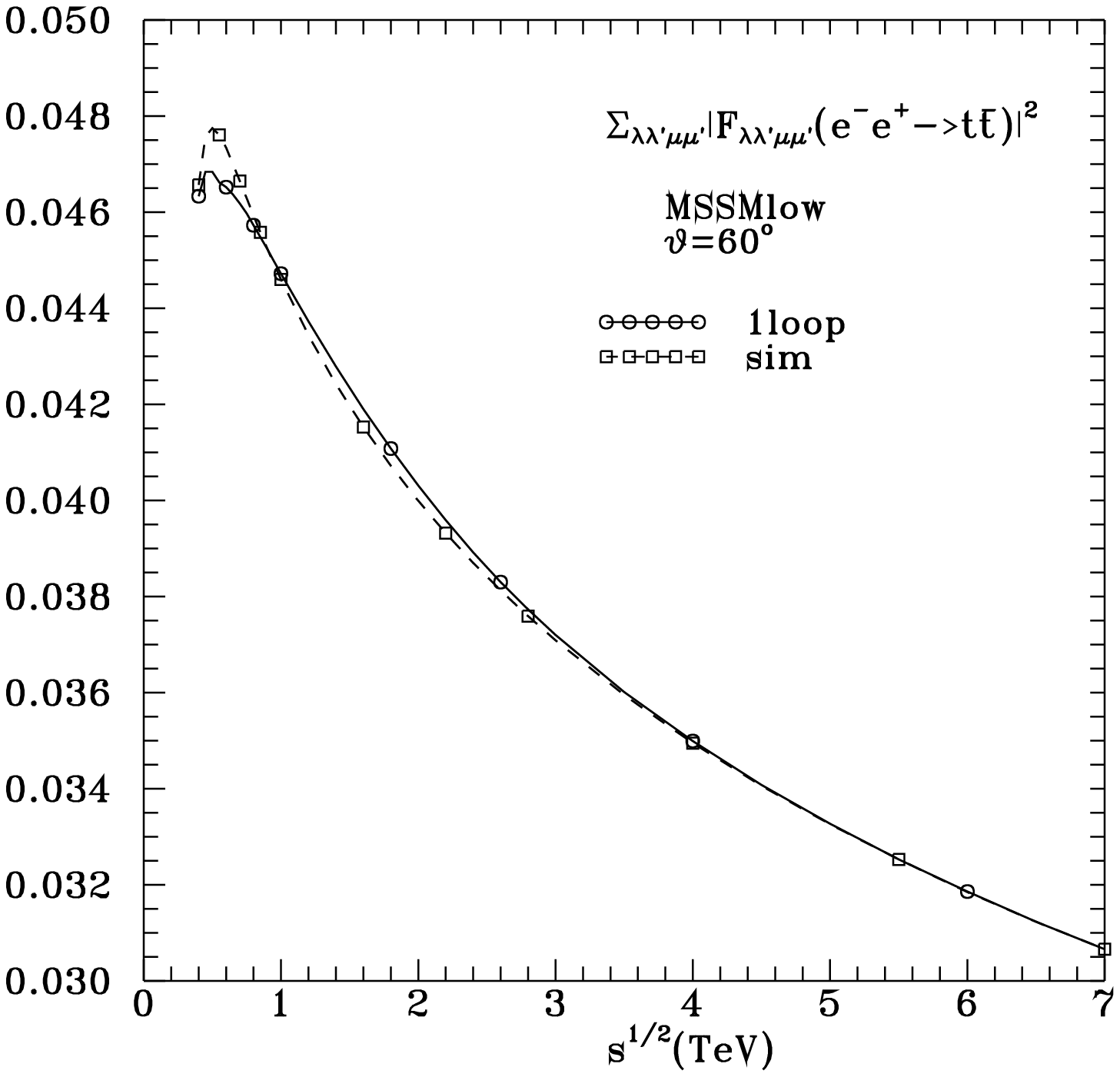, height=7.cm}\hspace{1.cm}
\epsfig{file=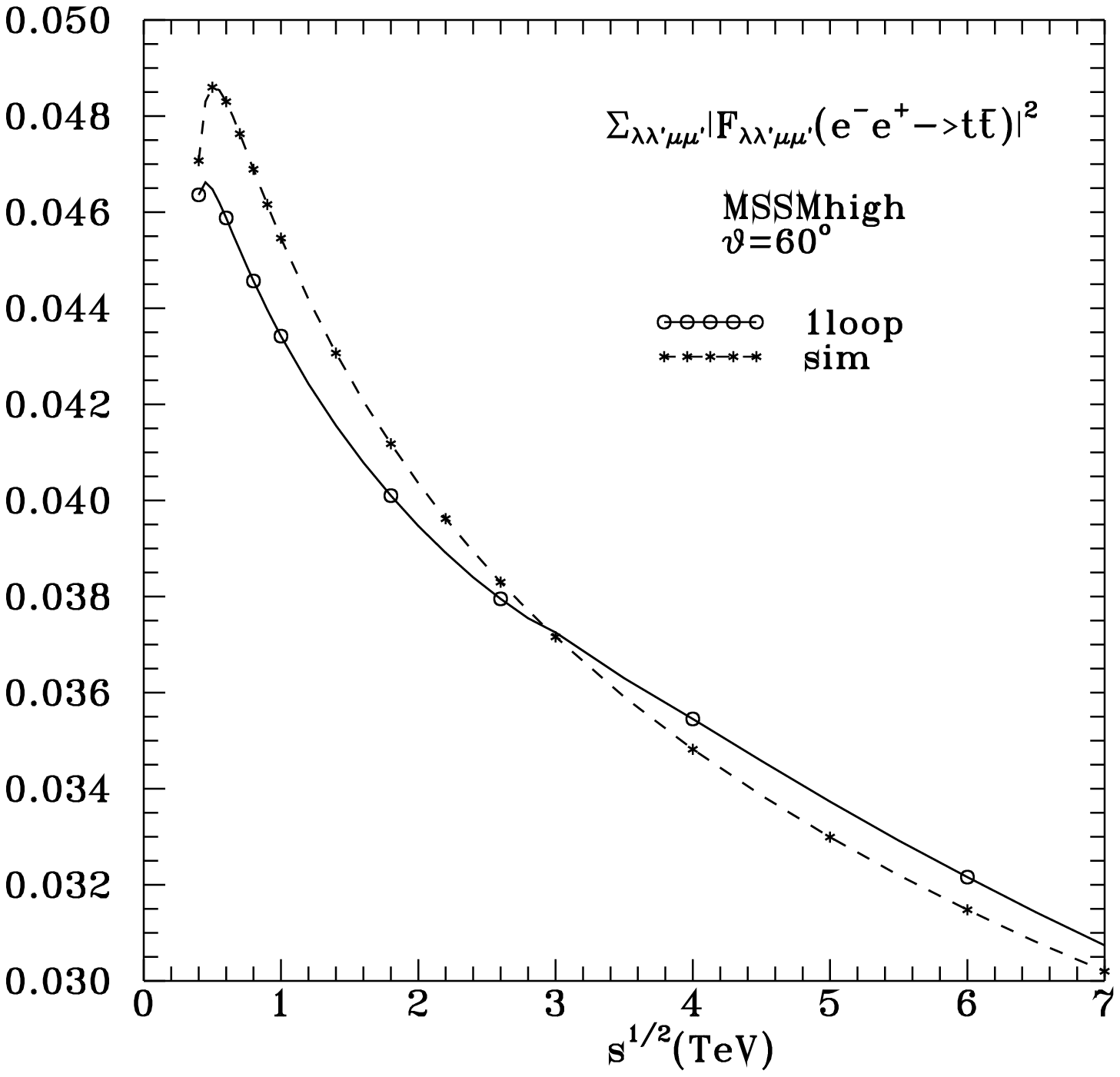,height=7.cm}
\]
\caption[1]{Energy dependence for the "dimensionless cross section" in
(\ref{sig-reduced}). Full lines describe
 the 1loop EW order results, while broken lines describe the  "sim" results  determined
  as stated just after (\ref{sig-reduced}).
 Models and panels as in Fig.\ref{HV-fig}.}
\label{sigma-e-fig}
\end{figure}

\begin{figure}[d]
\[
\vspace*{-0.5cm}
\hspace{-1.cm}
\epsfig{file=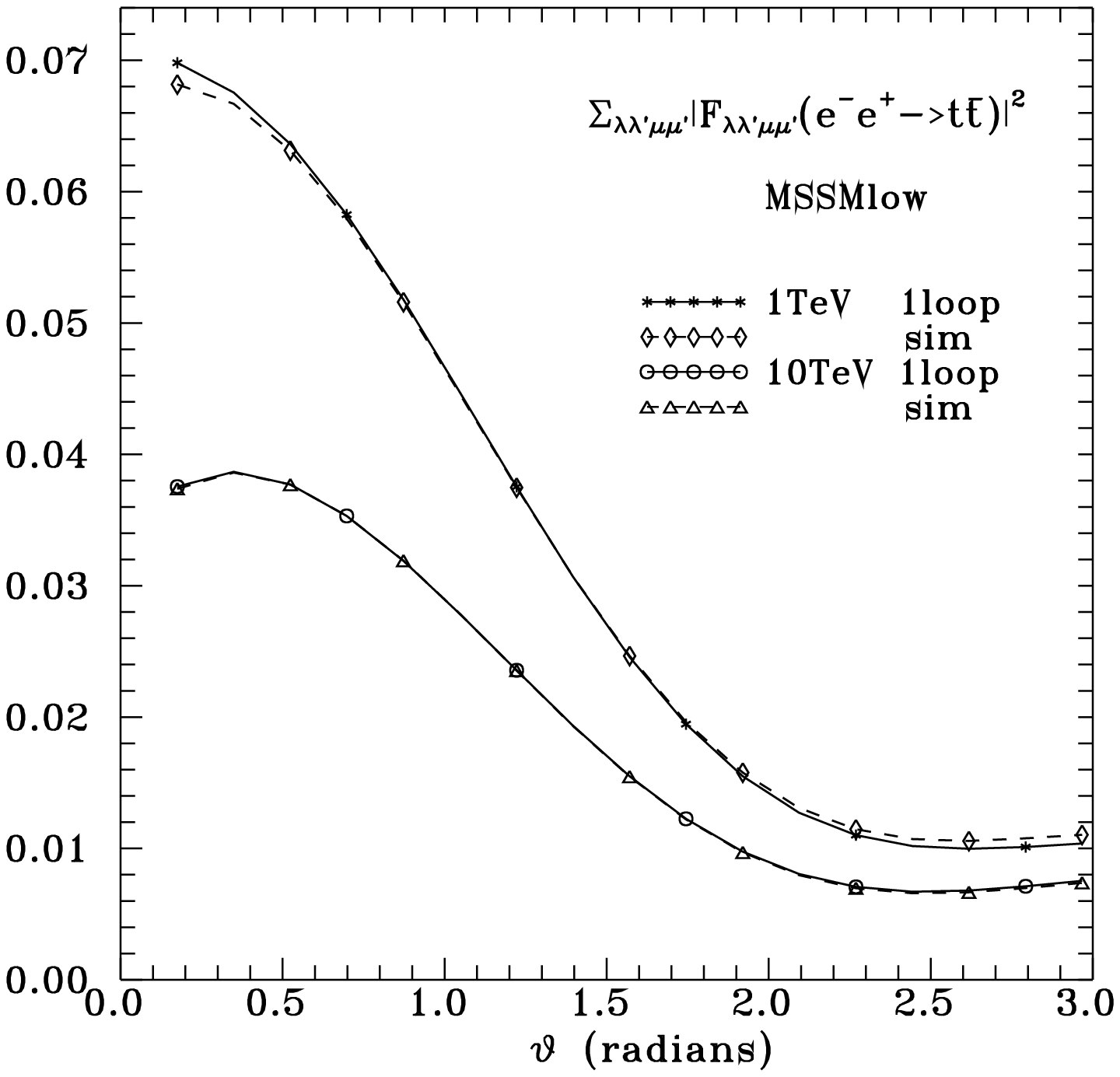, height=7.cm}\hspace{1.cm}
\epsfig{file=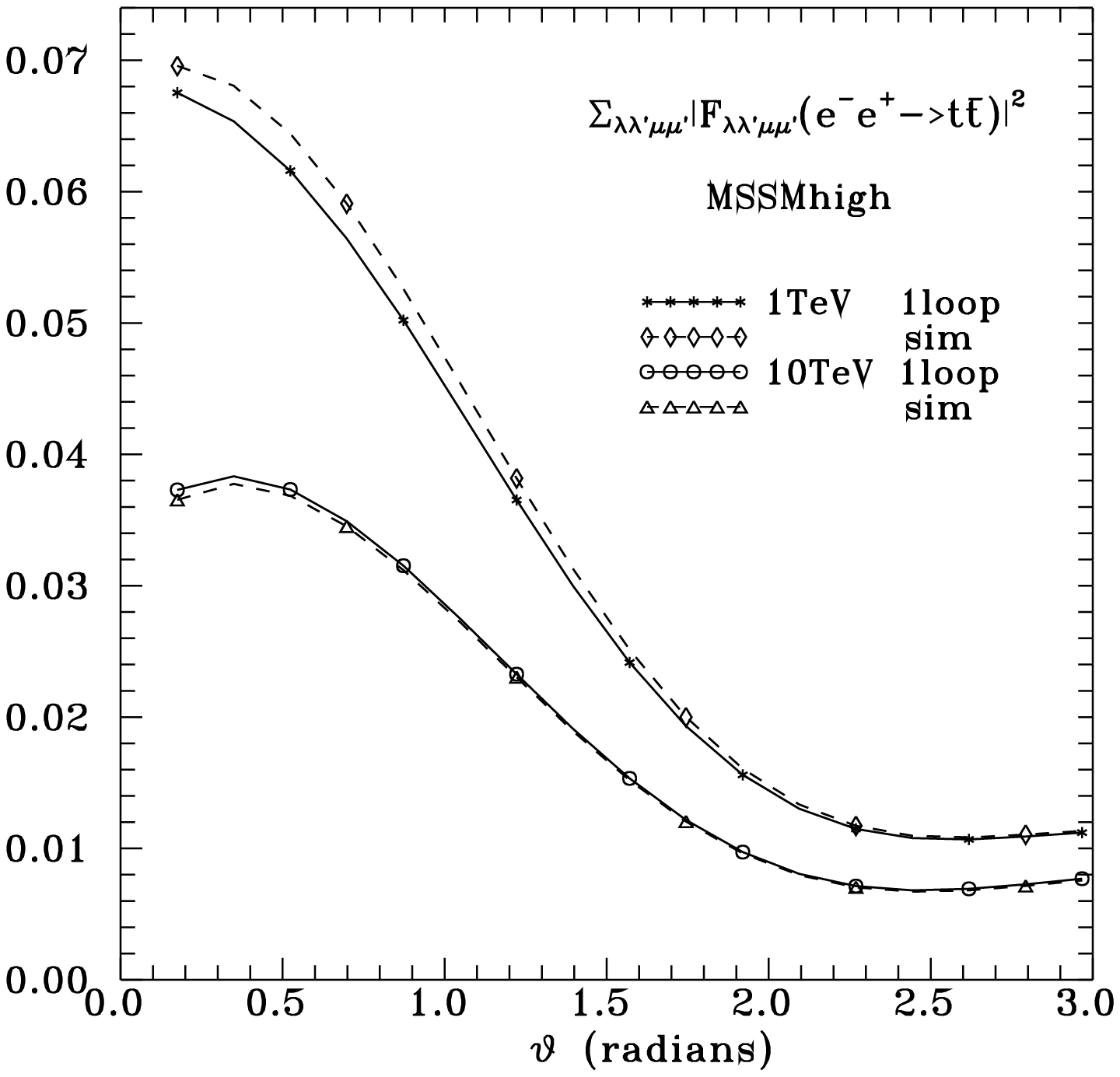,height=7.cm}
\]
\caption[1]{Angular dependence for the "dimensionless cross section" in
(\ref{sig-reduced}), at c.m. energies of 1  and 10 TeV. Full lines describe
 the  1loop EW order results, while broken lines describe the  "sim" results
 determined as stated  just after (\ref{sig-reduced}).
 Models and panels as in Fig.\ref{HV-fig}.}
\label{sigma-a-fig}
\end{figure}

Comparing Fig.\ref{HV-fig} and \ref{HC-amp-fig} we can also see
that for both MSSMhigh and MSSMlow at 1 TeV and  $\theta=60^o$, the HC amplitudes are already
dominating the HV ones; so that the HV contribution to the cross sections is at the 3\% level.
Varying the angle, changes relative individual contributions from various HC and  HV amplitudes;
but globally the ratio of their contributions to the cross section
remains at this level. Above 1 TeV of course, the HV contribution to the cross section
 falls quickly down.\\

These results have interesting theoretical and  predictive implications.

Theoretically they emphasize the elegance  of Supersymmetry,
where  the joint action of standard
and of spartner states, produces remarkable structures for the amplitudes
at high energy. More explicitly SUSY  forces all helicity violating 2-to-2 amplitudes
to vanish exactly at high energy \cite{heli1,heli2}; while it assigns
to the HC amplitudes at the 1loop EW order, very simple
and accurate expressions \cite{super}.

For a SUSY scale in the range of the above two benchmarks,
the predictive power of  the supersimple description
reaches the accuracy of  the percent level, at reasonable energies.
It can therefore be used to calculate
the values of physical  observables, keeping  the identification of the
important physical input clear. In other words, without relying
on enormous  codes, where the main  physical reason and the many  minor effects,
are thoroughly interwoven. For example, we have shown that the supersimplicity expressions
may be useful for immediately distinguished the MSSM  effects, from possible top-related
new  physics, beyond it.

\vspace*{1cm}

\renewcommand{\thesection}{A}
\renewcommand{\theequation}{A.\arabic{equation}}
\setcounter{equation}{0}

\section{Appendix: High energy   HC amplitudes}

Defining the momenta and helicities for the  $e^-e^+\to t \bar t $ process as in
(\ref{eett-process}), and the helicity amplitudes as in (\ref{F-amplitude}), the Born
contributions are given by
\bqa
F^{\rm Born}_{\lambda\lambda'\mu\mu'}
& = &\sum_{V=\gamma,Z}{1\over s-m^2_V}\bar u_t(p,\mu)\gamma^{\mu}(g^{L}_{Vt}P_L+g^{R}_{Vt}P_R)
v_t((p',\mu') \nonumber \\
& \cdot & \bar v_e(l',\lambda' ) \gamma_{\mu}(g^{L}_{Ve}P_L+g^{R}_{Ve}P_R)u_e(l,\lambda)~~,
\label{FBorn}
\eqa
where  $(l,l',p,p')$ denote the momenta and
$(\lambda, \lambda', \mu, \mu') $ the helicities, of the incoming
and outgoing particles, using the standard conventions \cite{JW}.
Neglecting all masses at high energies, the  Mandelstam variables are
\bqa
s&=& (l+l')^2=(p+p')^2 ~~ , \nonumber \\
t &= &(l-p)^2=-\frac{s}{2}(1-\cos\theta) ~~ , \nonumber \\
u &= &(l-p')^2=-\frac{s}{2}(1+\cos\theta) ~~ , \label{stu-variables}
\eqa
where   $\theta$ is  the c.m. scattering angle. Finally
\bqa
&& g^L_{\gamma e}=g^R_{\gamma e}=-e ~~,~~  g^L_{\gamma t}=g^R_{\gamma t}={2e\over3}
~~, \nonumber \\
&& g^L_{Z e}={e(-1+2s^2_W)\over2s_Wc_W} ~~,~~  g^R_{Z e}={es_W\over c_W}
~~,~~ \nonumber  \\
&& g^L_{Z t}={e(3-4s^2_W)\over6s_Wc_W} ~~,~~ g^R_{Z t}={-2es_W\over3c_W} ~~ \label{gammaZ-coup}
\eqa
denote  the usual SM couplings.

Neglecting $m_e$, there exist only four independent HC amplitudes
$F_{-+-+}$, $F_{+-+-}$, $F_{-++-}$, $F_{+--+}$
 to be considered, whose  Born contributions at high energies are
\bqa
F^{\rm Born}_{-+-+}
\simeq e^2 \left ({2u\over s}\right )\left ({-3+2s^2_W\over12s^2_Wc^2_W}\right ) ~&,&~
F^{\rm Born}_{+-+-}\simeq e^2 \left ({2u\over s}\right)\left  ({-2\over3c^2_W}\right ) ~~,
\nonumber \\
F^{\rm Born}_{-++-}\simeq e^2\left ({2t\over s}\right )\left ({-1\over3c^2_W}\right ) ~&,&~
F^{\rm Born}_{+--+}\simeq e^2\left ({2t\over s}\right )\left ({-1\over6c^2_W}\right ) ~~.
\label{F-Born-HC}
\eqa\\

\subsection{The supersimple  augmented Sudakov amplitudes.}

The  high energy supersimple  expressions for  the Sudakov part
 of the HC amplitudes, at the  1loop EW order,  are
\bqa
&& F^{\rm Sud}_{-+-+}\simeq {2u\alpha^2\over s} \Bigg \{
 {(9-12s^2_W+4s^4_W)\over144s^4_Wc^4_W}\Big [{(u-t)\over u}(\ln^2r_{ts}+\pi^2)
-2\ln^2 t_Z +2\ln^2 u_Z\Big ]
\nonumber\\
&&-{(45-84s^2_W+40s^4_W)\over72s^4_Wc^4_W}(\ln^2r_{us}+\pi^2)
+{1\over2s^4_W} \Big [\ln^2 u_W+2L_{eW\nu}+2L_{tWb}
\nonumber\\
&&
-{1\over2}\overline{\ln(s_{W\nu})}-2\overline{\ln(s_{WW})}
-{1\over2}\overline{\ln(s_{Wb})} \Big ]
\nonumber\\
&&-{(3-4s^2_W)\over24s^4_Wc^2_W}\Big [2\ln^2 s_W+4L_{eW\nu}
-3\overline{\ln(s_{W\nu})}+4L_{tZt}
-3\overline{\ln(s_{Zt})} \Big ]
\nonumber\\
&&-{(-3+2s^2_W)\over48s^4_Wc^4_W} \Big [\ln^2 s_Z+4L_{eZe}
-3\overline{\ln(s_{Ze})}\Big ]
\nonumber\\
&&-{(-27+42s^2_W-16s^4_W)\over 432s^4_Wc^4_W} \Big [\ln^2 s_Z +4L_{tZt}
-3\overline{\ln(s_{Zt})}\Big ]
\nonumber\\
&&+{(3-2s^2_W)\over48s^4_Wc^4_W}
\sum_i \Big [|Z^N_{1i}s_W+Z^N_{2i}c_W|^2\overline{\ln(s_{\tchi^0_i
\tilde{e}_L})}+{|Z^N_{1i}s_W+3Z^N_{2i}c_W|^2\over9}\overline{\ln(s_{\tchi^0_i \tilde{t}_L})}\Big ]
\nonumber\\
&&+ {(3-5s^2_W+2s^4_W)\over24s^4_Wc^4_W} \sum_i|Z^+_{1i}|^2
\overline{\ln(s_{\tchi^+_i \tilde{\nu}_L})}
+{(3-2s^2_W) \over24s^4_Wc^2_W} \sum_i|Z^-_{1i}|^2\overline{\ln(s_{\tchi^+_i \tilde{b}_L})}
\nonumber\\
&&+{(3-2s^2_W)\over24s^2_Wc^2_W}\Big [{m^2_t\over 2 s^2_Wm^2_W\sin^2\beta} \sum_i|Z^N_{4i}|^2
\overline{\ln(s_{\tchi^0_i \tilde{t}_R})}+{m^2_b\over 2s^2_Wm^2_W\cos^2\beta}\sum_i|Z^+_{2i}|
\overline{\ln(s_{\tchi^+_i \tilde{b}_R})}\Big ]
\nonumber\\
&&+{3-2s^2_W\over48s^2_Wc^2_W}\Big [ {m^2_t\over 2s^2_Wm^2_W}
\Big ({\sin^2\alpha\over\sin^2\beta}\, \overline{\ln(s_{tH^0} )}
+ {\cos^2\alpha\over\sin^2\beta}\,
\overline{\ln(s_{th^0})}+{\cos^2\beta\over\sin^2\beta}\,
\overline{\ln(s_{tA^0})}+\overline{\ln(s_{tG^0})}\Big )
\nonumber\\
&& + {m^2_b\over s^2_Wm^2_W}
\Big (\tan^2\beta\, \overline{\ln(s_{bH^+})}+\overline{\ln(s_{bG^+})}\Big )\Big ]\Bigg \}
~~, \label{sim-FSudmpmp}
\eqa
\bqa
&& F^{\rm Sud}_{+-+-}\simeq  {2u\alpha^2\over s} \Bigg \{{4\over9c^4_W}
\Big [{(u-t)\over u}(\ln^2r_{ts}+\pi^2)
-2(\ln^2r_{us}+\pi^2+\ln^2 t_Z -\ln^2 u_Z)\Big ]
\nonumber\\
&&+ {2\over3c^4_W} \Big [\ln^2 s_Z +4L_{eZe} -3\overline{\ln(s_{Ze})}
+ {4\over9} [\ln^2 s_Z +4L_{tZt} -3\overline{\ln(s_{Zt})}] \Big ]
\nonumber\\
&&+{2\over3c^4_W} \sum_i \Big [|Z^N_{1i}|^2\overline{\ln(s_{\tchi^0_i \tilde{e}_R})}
+{4\over9}|Z^N_{1i}|^2\overline{\ln(s_{\tchi^0_i \tilde{t}_R})} \Big ]
\nonumber\\
&&+ {m^2_t\over 6s^2_Wc^2_Wm^2_W \sin^2\beta } \sum_i \Big [|Z^N_{4i}|^2
\overline{\ln(s_{\tchi^0_i \tilde{t}_L})}+|Z^+_{2i}|^2
\overline{\ln(s_{\tchi^+_i \tilde{b}_L})} \Big ]
\nonumber\\
&&+ {m^2_t\over 12s^2_Wc^2_Wm^2_W} \Big [{\sin^2\alpha\over\sin^2\beta}
\overline{\ln(s_{tH^0})}+ {\cos^2\alpha\over\sin^2\beta}
\overline{\ln(s_{th^0})}+{\cos^2\beta\over\sin^2\beta}
\overline{\ln(s_{tA^0})}+\overline{\ln(s_{tG^0})}
\nonumber\\
&& +2\cot^2\beta\overline{\ln(s_{bH^+})}+2\overline{\ln(s_{bG^+})} \Big ]\Bigg \}
~~, \label{sim-FSudpmpm}
\eqa
\bqa
&& F^{\rm Sud}_{-++-}\simeq {-2t\alpha^2\over s} \Bigg \{
{1\over9c^4_W} \Big [-2(\ln^2r_{ts}+\pi^2)
+{(t-u)\over t}(\ln^2r_{us}+\pi^2)+2\ln^2 t_Z -2ln^2 u_Z \Big ]
\nonumber\\
&&- {1\over12s^2_Wc^4_W} \Big [\ln^2 s_Z +4L_{eZe}
-3\overline{\ln(s_{Ze})}+{16s^2_W\over9}\Big (\ln^2 s_Z+4L_{tZt}
-3\overline{\ln(s_{Zt})} \Big) \Big ]
\nonumber\\
&&- {1\over6s^2_Wc^2_W} \Big [\ln^2 s_W +4L_{eW\nu}-3\overline{\ln(s_{W\nu})} \Big ]
\nonumber\\
&&-{1\over12s^2_Wc^4_W} \sum_i\Big [\Big |Z^N_{1i}s_W+Z^N_{2i}c_W \Big |^2
\overline{\ln(s_{\tchi^0_i
\tilde{e}_L})}+{16s^2_W\over9}|Z^N_{1i}|^2\overline{\ln(s_{\tchi^0_i\tilde{t}_R})} \Big ]
\nonumber\\
&&-{1\over6s^2_Wc^2_W}\sum_i|Z^+_{1i}|^2\overline{\ln(s_{\tchi^+_i\tilde{\nu}_L})}
- {m^2_t\over 12s^2_Wc^2_Wm^2_W \sin^2\beta } \sum_i \Big [|Z^N_{4i}|^2
\overline{\ln(s_{\tchi^0_i \tilde{t}_L})}+|Z^+_{2i}|
\overline{\ln(s_{\tchi^+_i \tilde{b}_L})} \Big ]
\nonumber\\
&&- {m^2_t\over 24s^2_Wc^2_Wm^2_W} \Big [{\sin^2\alpha\over\sin^2\beta}
\overline{\ln(s_{tH^0})}+ {\cos^2\alpha\over\sin^2\beta}
\overline{\ln(s_{th^0})}+{\cos^2\beta\over\sin^2\beta}
\overline{\ln(s_{tA^0})}+\overline{\ln(s_{tG^0})}
\nonumber\\
&& +2\cot^2\beta\overline{\ln(s_{bH^+})}+2\overline{\ln(s_{bG^+})}) \Big ] \Bigg \}
~~, \label{sim-FSudmppm}
\eqa
\bqa
&& F^{\rm Sud}_{+--+}\simeq {-2t\alpha^2\over s}\Bigg \{
{1\over36c^4_W} \Big [-2(\ln^2r_{ts}+\pi^2)
+{(t-u)\over t} (\ln^2r_{us}+\pi^2)+2\ln^2t_Z -2\ln^2 u_Z \Big ]
\nonumber\\
&&-{1\over6c^4_W} \Big [\ln^2 s_Z +4L_{eZe} -3\overline{\ln(s_{Ze})} \Big ]
- {(9-8s^2_W) \over216s^2_Wc^4_W} \Big [\ln^2 s_Z +4L_{tZt} -3\overline{\ln(s_{Zt})} \Big ]
\nonumber\\
&&-{1\over12s^2_Wc^2_W} \Big [\ln^2 s_W+4L_{tWb}-3\overline{\ln(s_{Wb})} \Big ]
\nonumber\\
&&- {1\over6c^4_W} \sum_i\Big [|Z^N_{1i}|^2\overline{\ln(s_{\tchi^0_i \tilde{e}_R})}
+{1\over36s^2_W}|Z^N_{1i}s_W+3Z^N_{2i}c_W|^2\overline{\ln(s_{\tchi^0_i \tilde{t}_L})}\Big ]
\nonumber\\
&&- {1\over12s^2_Wc^2_W} \sum_i \Big [|Z^-_{1i}|^2
\overline{\ln(s_{\tchi^+_i \tilde{b}_L})}+{m^2_t|Z^N_{4i}|^2 \over 2m^2_W \sin^2\beta }
\overline{\ln(s_{\tchi^0_i \tilde{t}_R})}+{ m^2_b |Z^+_{2i}|^2\over 2m^2_W\cos^2\beta }
\overline{\ln(s_{\tchi^+_i \tilde{b}_R})} \Big ]
\nonumber\\
&&- {m^2_t\over 48s^2_Wc^2_Wm^2_W} \Big [{\sin^2\alpha\over\sin^2\beta}
\overline{\ln(s_{tH^0})}+ {\cos^2\alpha\over\sin^2\beta}
\overline{\ln(s_{th^0})}+{\cos^2\beta\over\sin^2\beta}
\overline{\ln(s_{tA^0})}+\overline{\ln(s_{tG^0})} \Big ]
\nonumber\\
&&- {m^2_b\over 24s^2_Wc^2_Wm^2_W} \Big [
\tan^2\beta\overline{\ln(s_{bH^+})}+\overline{\ln(s_{bG^+})} \Big ] \Bigg \}
~~ \label{sim-FSudpmmp}
\eqa
where the
chargino and neutralino mixing matrices $(Z^N, Z^+, Z^-)$  are as in \cite{Rosiek},

Note that all high energy supersimple expressions (\ref{sim-FSudmpmp}-\ref{sim-FSudpmmp})
are solely expressed as linear combinations of the
forms (\ref{Sud-ln2}, \ref{Sud-ln}, \ref{r-form}), with the coefficients
of  (\ref{Sud-ln2}, \ref{Sud-ln}) being
constants satisfying  the general constraints \cite{MSSMrules1,MSSMrules2,MSSMrules3,MSSMrules4}.
The coefficients of the forms (\ref{r-form}) though,  may also involve ratios
of Mandelstam variables \cite{super}. No additional constants appear in
(\ref{sim-FSudmpmp}-\ref{sim-FSudpmmp}); i.e. there are
no additional residual terms  in them.

We also remark that  the pinch contribution which  only affects $F_{-+-+}$,
has been put in  $F^{\rm s.e.}_{-+-+}$, as discussed in Sect 2.

Notice also that for the $e^-e^+\to t \bar t$, the structure
of  (\ref{Sud-ln2},\ref{LaVc-term},\ref{lambda-function}) implies that
\bq
\ln^2 t_Z - \ln^2 u_Z = \overline{\ln^2 t_Z} - \overline{\ln^2 u_Z} ~~,
\label{ln2-tZ-uZ-eq}
\eq
so that all  $\ln^2(x_V)$  terms  in (\ref{sim-FSudmpmp}-\ref{sim-FSudpmmp})
with $(x=s,t,u)$, are  consistent with  the form (\ref{Sud-ln2}).

Moreover, since  we are using a Feynman-t'Hooft gauge,
the masses of the charged and neutral Goldstone bosons
(whenever they appear in the equations above)
are taken as  $m_W$ and $m_Z$ respectively.

Finally,  (\ref{sim-FSudmpmp}-\ref{sim-FSudpmmp}) clearly indicate
that the Yukawa interactions do respect the supersimplicity structure. \\

\subsection{The supersimple augmented RG amplitudes.}

At high energies the $\gamma\gamma, \gamma Z, ZZ$ renormalized
self-energy contribution to the
four HC helicity amplitudes, together with the pinch contribution, are
\bqa
&& F^{\rm s.e.}_{-+-+} \simeq -{2u\over s^2}\sum_{V,V'} \hat{\Sigma}_{VV'}(s)g^L_{Ve}g^L_{V't}
+ \frac{\alpha^2}{\sw^4 }\, \overline{\ln s_{WW}} ~~,
\label{sim-Fsempmp} \\
&& F^{\rm s.e.}_{+-+-} \simeq -{2u\over s^2}\sum_{V,V'} \hat{\Sigma}_{VV'}(s)g^R_{Ve}g^R_{V't}
~~, \label{sim-Fsepmpm}  \\
&& F^{\rm s.e.}_{-++-}\simeq -{2t\over s^2}\sum_{V,V'} \hat{\Sigma}_{VV'}(s)g^L_{Ve}g^R_{V't}
~~,  \label{sim-Fsemppm}  \\
&& F^{\rm s.e.}_{+--+}\simeq -{2t\over s^2} \sum_{V,V'} \hat{\Sigma}_{VV'}(s)g^R_{Ve}g^L_{V't}
~~,  \label{sim-Fsepmmp}
\eqa
where $V$ and $V'$ run over $\gamma$ and  $Z$, and the coupling constants are
given in (\ref{gammaZ-coup}). The last term in (\ref{sim-Fsempmp}) is the aforementioned
pinch contribution (\ref{pinch-term}).

We next discuss the renormalized gauge self-energy functions $\hat{\Sigma}_{VV'}(s)$.
In the on-shell scheme we have (for details and notations see \cite{OS})
\bqa
&& \hat{\Sigma}_{\gamma\gamma}(s)=\Sigma_{\gamma\gamma}(s)+s\delta Z^{\gamma}_2
~~, \nonumber\\
&&\hat{\Sigma}_{ZZ}(s)=\Sigma_{ZZ}(s)-\delta m^2_Z+(s-m^2_Z)\delta Z^{Z}_2
~~, \nonumber\\
&&\hat{\Sigma}_{\gamma Z}(s)=\Sigma_{\gamma Z}(s)+s\delta Z^{\gamma Z}_2
+ m^2_Z(\delta Z^{\gamma Z}_1-\delta Z^{\gamma Z}_2)~~, \label{hatSigmaVV}
\eqa
where
\bqa
&&\delta Z^{\gamma}_2=-~\Sigma^{'}_{\gamma\gamma}(0)~~~,~~~
\delta Z^{\gamma}_1=-~\Sigma^{'}_{\gamma\gamma}(0)
+{s_W\over c_W}{\Sigma_{\gamma Z}(0)\over  m^2_Z}
~~, \nonumber\\
&&\delta Z^{Z}_2=-~\Sigma^{'}_{\gamma\gamma}(0)
+2{c^2_W-s^2_W\over s_Wc_W}{\Sigma_{\gamma Z}(0)\over  m^2_Z}
+{c^2_W-s^2_W\over s^2_W}\left ({\delta m^2_Z\over m^2_Z}-{\delta m^2_W\over m^2_W}\right )
~~, \nonumber\\
&&\delta Z^{Z}_1=-~\Sigma^{'}_{\gamma\gamma}(0)
+{3c^2_W-2s^2_W\over s_Wc_W}{\Sigma_{\gamma Z}(0)\over  m^2_Z}
+{c^2_W-s^2_W\over s^2_W}\left ({\delta m^2_Z\over m^2_Z}-{\delta m^2_W\over m^2_W}\right )
~~, \nonumber \\
&& \delta Z^{\gamma Z}_i={c_Ws_W\over c^2_W-s^2_W}(\delta Z^{Z}_i-\delta Z^{\gamma}_i)
~~, \nonumber \\
&& \delta M^2_W=Re\Sigma_{WW}(M^2_W)~~~,~~~\delta M^2_Z=Re\Sigma_{ZZ}(M^2_Z)
 ~~~ . \label{gauge-ct}
\eqa\\

At high energies, (\ref{hatSigmaVV}) become
\bqa
&& \hat{\Sigma}_{\gamma\gamma}(s) \simeq \Sigma_{\gamma\gamma}(s)+s\delta Z^{\gamma}_2
~~, \nonumber\\
&&\hat{\Sigma}_{ZZ}(s)\simeq  \Sigma_{ZZ}(s)+s\delta Z^{Z}_2
~~,  \nonumber\\
&&\hat{\Sigma}_{\gamma Z}(s) \simeq  \Sigma_{\gamma Z}(s)+s\delta Z^{\gamma Z}_2
~~. \label{asym-SigmaVV}
\eqa
Using then the definitions (\ref{Sud-ln},\ref{b0mZ}),
  we obtain  the  supersimple high energy expressions
\bqa
&& \hat{\Sigma}_{\gamma\gamma}(s) \simeq  {s\alpha\over\pi}
\Bigg \{ {3\over4}\overline{\ln(s_{W^+W^-})}
-~{1\over12}\overline{\ln(s_{H^+H^-})}
\nonumber\\
&&-\sum_f N^f_cQ^2_f {1\over3}\Big [
\overline{\ln(s_{f\bar f})}
+{1\over4}\sum_i \overline{\ln(s_{\tilde{f}_i\bar {\tilde{f}_i}})} \Big ]
-{1\over3}\sum_{\tchi_j} \overline{\ln(s_{\tchi^+_j\tchi^-_j})}\Bigg \}
~,\label{sim-Sigma-gamgam}
\eqa
\bqa
&& \hat{\Sigma}_{\gamma Z }(s)
\simeq  -\frac{s\alpha}{\pi} \Bigg \{
\frac{\cos(2\theta_W)}{24s_Wc_W}\, \Big [ \overline{\ln(s_{H^+H^-})}
+\overline{\ln(s_{W^+W^-})}\Big ]
 -\frac{5c_W}{6s_W}\overline{\ln(s_{W^+W^-})}
 \nonumber \\
&&  + \sum_f N_c^f Q_f{v_f\over3}\overline{\ln(s_{f\bar f}})
+\frac{1}{12 s_Wc_W}
\sum_f N_c^f Q_f \Big \{ (I_3^f \csf^2-Q_f s^2_W)
\overline{\ln(s_{\tilde f_1 \tilde f_1}})
\nonumber \\
&& + (I_3^f \ssf^2-Q_f s^2_W)
\overline{\ln(s_{\tilde f_2 \tilde f_2}}) \Big \}
+ \frac{1}{12 s_Wc_W}
\sum_{j=1}^2 (O_{jj}^{ZL}+O_{jj}^{ZR})
\overline{\ln(s_{ \tchi_j^+ \tchi_j^-}}) \Bigg \}
\nonumber \\
&&+s{c_W\over s_W}\overline{\Delta\rho}
~,\label{sim-Sigma-gamZ}
\eqa
\bqa
&& \hat{\Sigma}_{ZZ}(s) \simeq \frac{s\alpha}{\pi}\Bigg \{ {1\over4s^2_Wc^2_W}\Big [
-{\sin^2(\beta-\alpha)\over12}\Big (\overline{\ln(s_{hZ})}
+\overline{\ln(s_{H^0A^0})}\Big )
\nonumber \\
&& -{\cos^2(\beta-\alpha)\over12}\Big (\overline{\ln(s_{H^0Z})}
+\overline{\ln(s_{hA^0})}\Big ) \Big ]
\nonumber \\
&&  -{cos^2(2\theta_W)\over12}\overline{\ln(s_{H^+H^-})}
+\Big ({10\over3} \cw^4-{\cos^2(2\theta_W)\over12}\Big )\overline{\ln(s_{W^+W^-})}
\nonumber \\
&& - \sum_f N_c^f  \Big \{
{(v_f^2+a_f^2)\over3} \overline{\ln(s_{f\bar f}})
 - \frac{1}{48   s^2_Wc^2_W} \sum_f N_c^f  \Big \{
 4 [I_3^f \csf^2-Q_f s^2_W]^2\overline{\ln(s_{\tilde f_1 \tilde f_1}})
\nonumber \\
&&  +\ssf^2\csf^2(\overline{\ln(s_{\tilde f_1 \tilde f_2}})
 +\overline{\ln(s_{\tilde f_2 \tilde f_1}}))
+4[I_3^f \ssf^2-Q_f s^2_W]^2\overline{\ln(s_{\tilde f_2 \tilde f_2}}) \Big \}
 \nonumber \\
&& + \frac{1}{24 s^2_Wc^2_W} \Big [
\sum_{i,j=1}^4 O_{ji}^{0ZL}O_{ij}^{0ZL} \overline{\ln(s_{\tchi_i^0 \tchi_j^0}})
 \nonumber \\
&& - \sum_{i,j=1}^2 \Big ( O_{ij}^{ZL}O_{ji}^{ZL}+O_{ij}^{ZR}O_{ji}^{ZR} \Big )
 \overline{\ln(s_{\tchi_i^+ \tchi_j^-}}) \Big ]
 \Bigg \}
 +s{c^2_W-s^2_W\over s^2_W}\overline{\Delta\rho}
~~~,\label{sim-Sigma-ZZ}
\eqa
where $\tilde \theta_f $ denotes the $(\tilde f_L,\tilde f_R) $-sfermion mixing angle,
and
\bqa
&& O^{0ZL}_{ij}=O^{0ZL*}_{ji}=-O^{0ZR}_{ji}=-O^{0ZR*}_{ij}=
Z^{N*}_{4i}Z^{N}_{4j}-Z^{N*}_{3i}Z^{N}_{3j}~~,
\nonumber \\
&&  O^{ZL}_{ij}= Z^{+*}_{1i}Z^{+}_{1j}+\delta_{ij}(c^2_W-
s^2_W)~~, \nonumber \\
&&  O^{ZR}_{Zij}= Z^{-}_{1i}Z^{-*}_{1j}
+\delta_{ij}(c^2_W-s^2_W) ~~, \label{xhi-0-pm-elements}
\eqa
with  the  $(Z^N, Z^+, Z^-)$ matrices being as  in Appendix A.1.
 Finally
\bq
v_f=\frac{I_3^f-2 Q_f s^2_W}{2s_Wc_W} ~~~,~~~
a_f=\frac{I_3^f}{2s_Wc_W} ~~ , \label{vf-af}
\eq
with $I_3^f$ being  the third component of the isospin of the L-fermion or sfermion fields,
 and $Q_f$ the corresponding electric charge.
 In all cases, CP conserving SUSY couplings are assumed. \\

It is worth remarking, that the high energy  expressions
(\ref{sim-Fsempmp}-\ref{sim-Fsepmmp}) for the RG amplitudes, do respect
the supersimplicity structure. In this respect we note that in addition to the forms
(\ref{Sud-ln2}, \ref{Sud-ln}, \ref{r-form}), they  also contain   residual
constant contributions  given by $\overline{\Delta\rho}$
in  (\ref{sim-Sigma-gamZ},\ref{sim-Sigma-ZZ}) and  further discussed in Sect.3.2.   \\

\renewcommand{\thesection}{B}
\renewcommand{\theequation}{B.\arabic{equation}}
\setcounter{equation}{0}

\section{ Appendix:  Anomalous effective $Zt\bar t$ coupling and $Z'$ effects}

\begin{figure}[p]
\vspace*{0.1cm}
\[
\hspace{-1.cm}
\epsfig{file=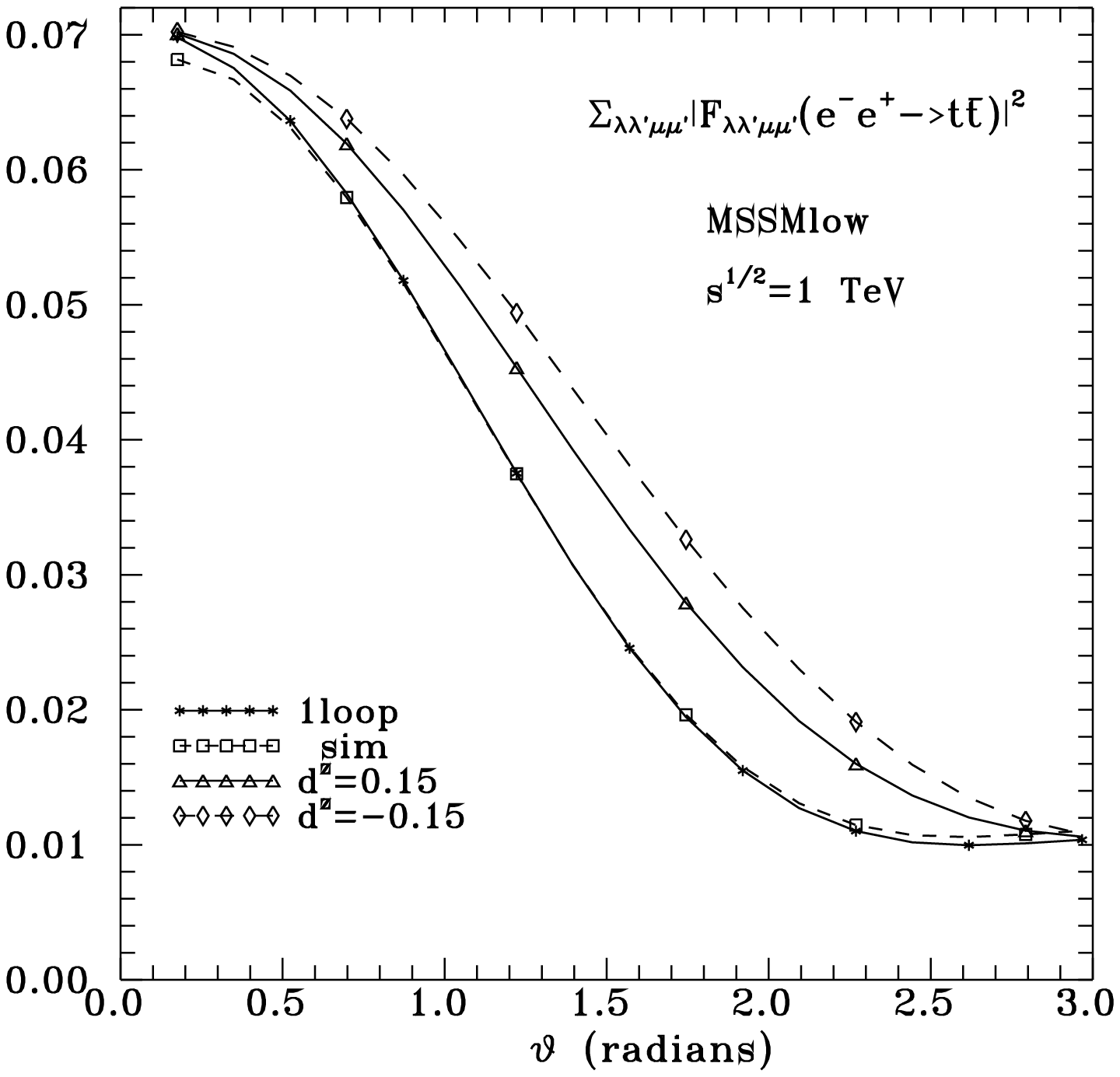, height=7.cm}\hspace{1.cm}
\epsfig{file=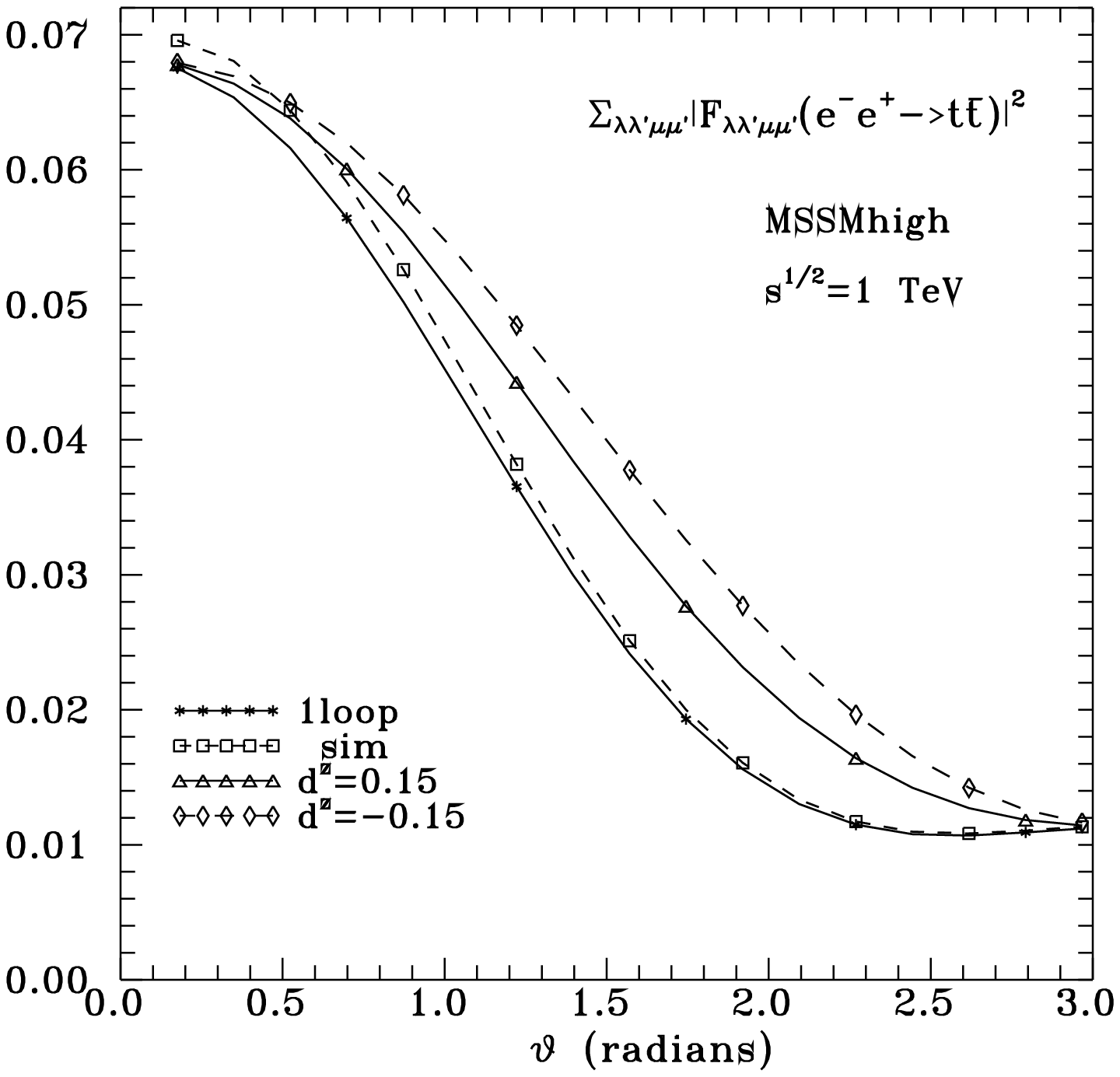,height=7.cm}
\]
\caption[1]{Angular dependence for the  "dimensionless cross section"   in
(\ref{sig-reduced}) and  the two MSSM benchmarks (\ref{model-MSSMlow},\ref{model-MSSMhigh}),
 at 1 TeV. Results including in addition the anomalous $Zt\bar t$ couplings  in (\ref{dZ}),
 are also shown. The   "sim" predictions  are described  just after (\ref{sig-reduced}).
Left panel corresponds to MSSMlow, and  right  panel  to
 MSSMhigh.}
\label{sigma-dz-fig}
\end{figure}

Here  we define the two new-physics  models, used for illustration
in Fig.\ref{sigma-dz-fig} and Table 1.

The first such model just contains the  additional  effective $Zt\bar t$ coupling
\bq
-ie{d^Z\over m_t} \epsilon^Z.(p-p')~~, \label{dZ}
\eq
where $p,p'$ denote the $t,\bar t$ momenta respectively; while
 $d^Z$ is an effective coupling, (which a priori could also be s-dependent).
Such an interaction leads to the additional helicity amplitudes
\bq
F^{d^Z}_{\lambda, \lambda',\mu,\mu'}=
-~{\lambda e^2d^{Z}s^{3/2} \over
2 m_ts^2_Wc^2_W(s-m^2_Z)}\left (1-{m^2_t\over s}\right )^2
\sin\theta \, \delta_{\mu,\mu'}
\left [g^{L}_{Ze}\delta_{\lambda,-{1\over2}}+g^{R}_{Ze}\delta_{\lambda,+{1\over2}} \right ]
~~, \label{F-dZ}
\eq
where   (\ref{eett-process}, \ref{F-amplitude}, \ref{gammaZ-coup}) are used.
As seen from (\ref{F-dZ}), $d^Z$-contributions only exist for the helicity violating
amplitudes  $F_{-+--}, ~F_{-+++}, ~F_{+---}, ~F_{+-++}$.\\

The second new-physics model used in Table 1, just contains a new vector boson $Z'$, with common,
purely vector or axial couplings, to all fermions. It is  described by the vertex
\bq
-ie \gamma^\nu \left [g^v_{Z'f} Z'_\nu(V)- g^a_{Z'f}\gamma_5 Z'_\nu(A) \right ]
~~. \label{Zprime}
\eq
In Table 1,  common couplings, for both
$f=e$ and  $f=t$ cases have been used, for purely vector (axial) couplings   chosen  as
$g^v_{Z'f}=1$  ($g^a_{Z'f}=1$).
The $Z'$-mass is taken as    $3{\rm TeV}$.

\end{document}